\documentclass{article}

\usepackage{arxiv}

\usepackage[backend=biber,
style=numeric,
sorting=none,
bibencoding=UTF-8
]{biblatex}
\usepackage[utf8]{inputenc} % allow utf-8 input
\usepackage[T1]{fontenc}    % use 8-bit T1 fonts
\usepackage{hyperref}       % hyperlinks
\usepackage{url}            % simple URL typesetting
\usepackage{booktabs}       % professional-quality tables
\usepackage{amsfonts}       % blackboard math symbols
\usepackage{nicefrac}       % compact symbols for 1/2, etc.
\usepackage{microtype}      % microtypography
\usepackage{lipsum}		% Can be removed after putting your text content
\usepackage{graphicx}
\usepackage{doi}
\usepackage{amsmath}
\usepackage{amssymb}
\usepackage{float}
\usepackage{bbold}
\usepackage{subcaption}
\usepackage{cleveref}
\usepackage{tikz}
\usepackage{tensor}

\crefformat{subfigure}{#2#1#3}
\crefrangeformat{subfigure}{#3#1#4--#5#2#6}
\crefmultiformat{subfigure}
  {#2#1#3}
  {,~#2#1#3}
  {,~#2#1#3}
  {,~#2#1#3}

\ExplSyntaxOn
\NewDocumentCommand{\onlyletter}{m}
 {
  \tl_set:Nx \l_tmpa_tl { #1 }
  \tl_item:Nn \l_tmpa_tl { -1 }
 }
\ExplSyntaxOff

\title{Optimal income crossover for two-class model using particle swarm optimization}

\author{Paulo H. dos Santos \\
	Departamento de Física\\
	Universidade Federal de Santa Catarina\\
	Florianópolis, 88040-900 \\
	\texttt{psantos.fsc@gmail.com} \\
	\And
	Igor D.S. Siciliani  \\
	Departamento de Física\\
	Universidade Federal de Santa Catarina\\
	Florianópolis, 88040-900 \\
	\texttt{igorschoeller@gmail.com} \\
	\And
	M.H.R. Tragtenberg  \\
	Departamento de Física\\
	Universidade Federal de Santa Catarina\\
	Florianópolis, 88040-900 \\
	\texttt{marcelo.tragtenberg@ufsc.br} \\
}

\renewcommand{\shorttitle}{Optimal Income Crossover}

\hypersetup{
pdftitle={A template for the arxiv style},
pdfsubject={q-bio.NC, q-bio.QM},
pdfauthor={David S.~Hippocampus, Elias D.~Striatum},
pdfkeywords={First keyword, Second keyword, More},
}

\begin{document}
\maketitle

\begin{abstract}
Personal income distribution may exhibit a two-class structure, such that the lower income class of the population (85-98\%) is described by exponential Boltzmann-Gibbs distribution, whereas the upper income class (15-2\%) has a Pareto power-law distribution. We propose a method, based on a theoretical and numerical optimization scheme, which allows us to determine the crossover income between the distributions, the temperature of the Boltzmann-Gibbs distribution and the Pareto index.
    
Using this method, the Brazilian income distribution data  provided by the National Household Sample Survey was studied.  The data was stratified into two dichotomies (sex/gender and color/race), so  the model was tested using different subsets along with accessing the economic differences between these groups. Lastly, we analyse the temporal evolution of the parameters of our model and the Gini coefficient discussing the implication on the Brazilian income inequality. 
    
To our knowledge, for the first time an optimization method is proposed in order to find a continuous two-class income distribution, which is able to delimit the boundaries of the two distributions. It also gives a measure of inequality which is a function that depends only on the Pareto index and the percentage of people in the high income region. It was found a temporal dynamics relation, that may be general, between the Pareto and the percentage of people described by the Pareto tail.
\end{abstract}

% keywords can be removed

\section{Introduction}
A long time ago, the economist Vilfredo Pareto identified a power-law behavior in the income distribution \cite{Pareto}. Pareto stated that the income probability density function describing this distribution is of the form

\begin{equation}\label{pareto}
    P(m)=b m^{-\alpha -1}  ,   
\end{equation}
where $m$ denotes the income, $\alpha$ is known as the Pareto index ranging between 1 and 3 and $b$ is a normalization constant. Later it was found that the Pareto law is suited for representing just the upper tail of the income distribution \cite{hill}.

The Pareto power-law was confirmed extensively on the upper income data from different countries \cite{EvIndia,EvJap}, and was also found to describe wealth distribution \cite{EvUS}. In this paper this high income region will be defined by the \textbf{top-percentage} indicator, which is the percentage of the population that follows Pareto behavior.

The Boltzmann-Gibbs distribution (BGD), in the classical kinetic theory, is the most probable energy distribution of a gas with elastic collisions in thermal equilibrium. It was later found to be very useful for modeling income distribution for the low and middle-income class, by setting energy to be the money of the agents \cite{Exp1,Exp2}. In a multi-agent simulations context, its asymptotic states were capable of displaying Boltzmann-Gibbs as well as Pareto statistical behaviors \cite{sim1,sim2}. Is worth mentioning that the most used distribution for this region is the log-normal distribution, however unlike the Boltzmann-Gibbs it is not a stationary distribution \cite{kalecki}.
Therefore, for income less than the crossover income, $m_c$, the distribution is given by

\begin{gather}\label{exp}
 \begin{cases}
      &P(m)=\frac{a}{T} \exp(\frac{-m}{T})  \\
      &m<m_c 
    \end{cases}    
    ,   
\end{gather}
where $a$ is a normalization parameter and $T$ is the "temperature" of the system. 

Consequently, the personal income distribution can be considered as a two-class structure, as the lower class of the population (85-98\%) is described by exponential BG distribution, whereas the upper class (15-2\%) follows a Pareto power-law distribution. The most used method to determine the crossover between these two regions is to use a fixed proportion for the Pareto tail based on a log-log graph, where the Pareto region will present a linear behavior \cite{2US,2BR,2RO}. However, this choice is rather subjective, and therefore, the crossover income determined is not optimal.

We propose in this paper a method to determine the total income distribution defined by parts, thus the crossover income can be established optimally. Firstly, it is defined a measure of goodness-of-fit statistics that will be minimized by a numerical algorithm called Particle Swarm Optimization (PSO) with limited-memory Broyden–Fletcher–Goldfarb–Shanno Bound (L-BFGS-B). We validate this method by studying the Brazilian income distribution using data from National Household Sample Survey (PNAD), an annual research available by the Brazilian Institute of Geography and Statistics (IBGE).

Among our findings, we highlight two of them, obtained from the study of the temporal evolution of the Brazilian income distribution. The first is the correlation between the Gini coefficient calculated with the data and the one calculated with the model. The second is the correlation between the Pareto index and the percentage of people that display the Pareto power-law behavior.

This paper is organized as follows. Section 2 explores the two-class complementary cumulative distribution function (CCDF) and its continuity. In section 3 is derived the relation between the parameters of our model and the Gini coefficient. Section 4 describes the L-BFGS-B Particle Swarm Optimization and justify our choice. We applied in section 5 the PSO optimizer fitting our model into the Brazilian income distribution data of the total population and the stratified population (sex/gender and color/race), performing a cross validation with a re-sampling technique. Section 6 explores the time evolution of the parameters of our model by fitting our model into the Brazilian income distribution in the years between 2001 and 2019. Conclusions are found in section 7.

\section{Two-class model for income distribution}

We can define a two-class model for a country income distribution using Eqs. (\ref{pareto},\ref{exp}) 

\begin{equation}\label{P}
    P(m)=\begin{cases}
          \frac{a}{T} e^{-m/T} \quad &m<m_c  \\
          b m^{-\alpha -1} \quad &m \geqslant m_c 
        \end{cases}    
        ,   
\end{equation}
equivalently CCDF
    
\begin{equation}\label{CMA}
 \int\limits ^{\infty }_{m} P( m') dm' \equiv \hat{C}(m)=
     \begin{cases}
           a\Bigl[\exp( -m/T) -\exp( -m_{c} /T)\Bigr] +\lambda \quad &m<m_c  \\
          \qquad \lambda\left(\frac{m}{m_c}\right)^{-\alpha } \quad &m \geqslant m_c
     \end{cases}    
        ,   
\end{equation}
where, by continuity, $\lambda=\tfrac{ b}{\alpha} m_c^{-\alpha}$ is the top percentage of income that follows the Pareto behavior, hence $\lambda$ is the top-percentage indicator mentioned before. The normalization give us

\begin{equation}\label{sigmall}
     \lambda=1-a(1-e^{-m_c/T})
\end{equation} 

Setting $a=1$ makes the Pareto distribution a correction to the exponential in the high income tail. This makes the parameters of the model more interpretable and easier to optimize. So Eqs. (\ref{CMA},\ref{sigmall}) become
%In section 5 we are going to test the model for a varying $a$, which we get values close to 1. 

\begin{equation}\label{CM}
     \hat{C}(m,\lambda,T,\alpha)=
     \begin{cases}
          \exp\Bigl( \frac{-m}{T}\Bigl)   \quad &m<m_c(\lambda,T)  \\[10pt] 
          \lambda\left(\frac{m}{m_c(\lambda,T)}\right)^{-\alpha } \quad &m \geqslant m_c(\lambda,T)
     \end{cases} ,   
   \end{equation}
and
\begin{equation}\label{sigmaM}
    m_c(\lambda,T)=T\log(\lambda^{-1}).
\end{equation} 

The CCDF in Eq. \ref{CM} is the two-class model by Yakovenko \cite{Exp3}. Two things to notice, first is the change in the parameters of the model that now are given by $(\lambda,T,\alpha)$, the second is the guarantee of the theoretical CCDF continuity, which is not always the case when the two distributions are fitted separately.

\begin{figure}[H]
    \centering
    \centerline{\includegraphics[width=0.7\linewidth]{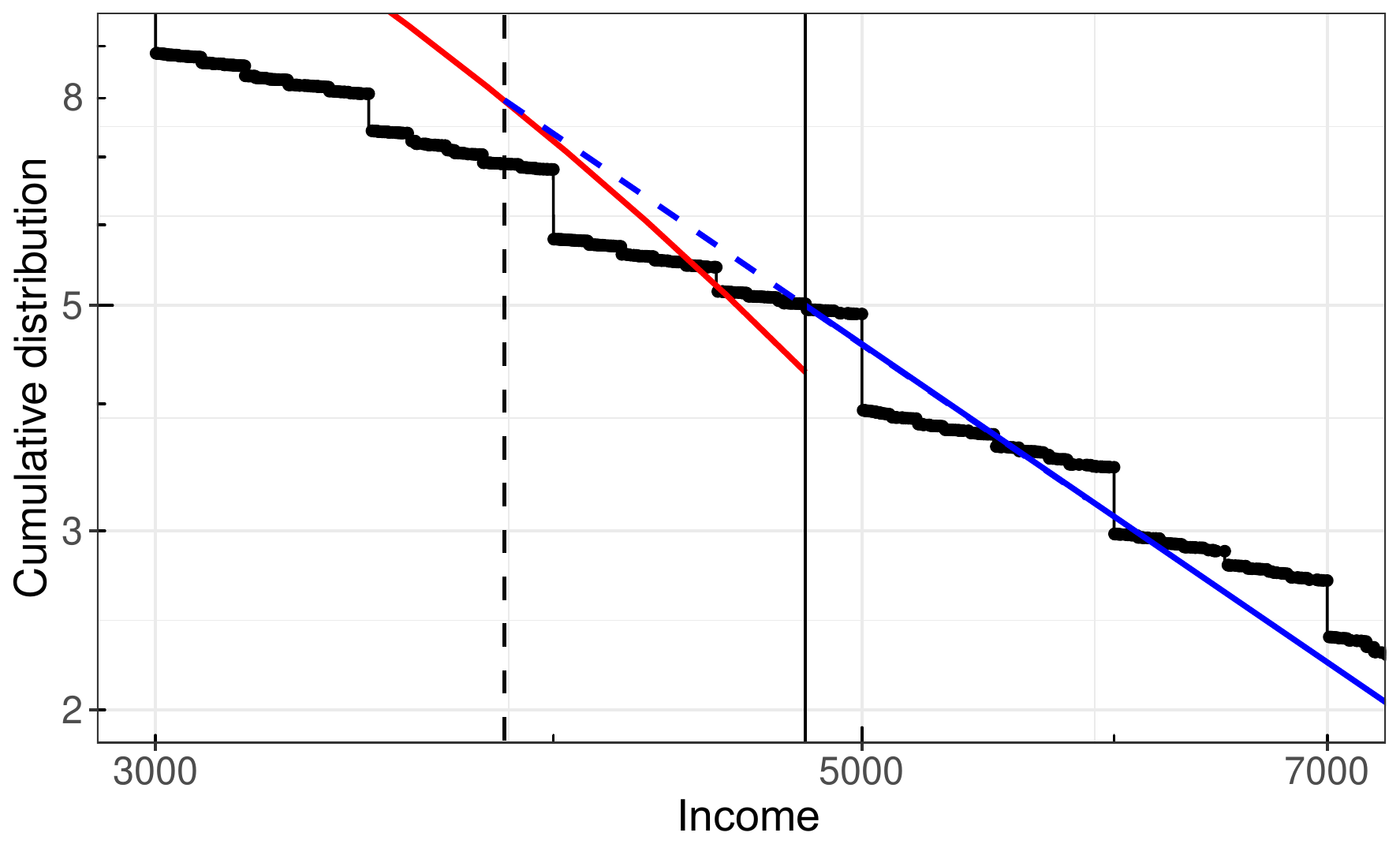}}
    \caption{\label{fig:extra} Cumulative probability distribution of income on a log–log scale. The black points represent the cumulative distribution of the data and the solid lines correspond to the fitted model described in Eq. \ref{CM}. The red curve obeys Boltzmann–Gibbs distribution, the power-law distribution is characterized by the blue curve and the dashed blue line is its extrapolation. The income crossover is represented by the vertical dashed line, whereas the Pareto threshold by the vertical solid line. }
\end{figure}

In previous methods of the two-class model, the $m_c$ is set by taking the intersection between the exponential fit (with $T=\langle m\rangle$) for the $98-85\%$ poorer people and a power-law fit in the $2-15\%$ richer region determined by the Pareto income threshold, sometimes needing an extrapolation as shown in Fig \ref{fig:extra}. Notice that the $m_c$ will not always be equal to this threshold, and, as stated before, CCDF will not be continuous in all cases.
The most commonly used method to determine the power-law income threshold is to plot the CCDF in a log-log scale and see where the behavior is linear. A more robust option is to use a goodness-of-fit function to optimize the threshold income \cite{Opthreshold}.  

In this paper we are going to determine the model parameters in Eq. \ref{CM} with a hybrid version of a numerical optimizer called Particle Swarm Optimizer. In order to achieve this, starting with a given set of income values $\{m_i\}$,  our method  will predict a vector of parameters $\{\lambda_p,T_p,\alpha_p\}$, of the income distribution $\hat{C}(m,\lambda_p,T_p,\alpha_p)$ given by equation (\ref{CM}). With this function we are going to be able to minimize the root mean squared logarithmic error (RMSLE) applying a hybrid approach of PSO. This method will display continuity and the equality between the Pareto income threshold and $m_c$.

\section{Gini coefficient} \label{secgini}

The Gini Coefficient, the most popular measure of income inequality, is derived from the Lorenz curve.

\begin{figure}[H]
    \centering
    \centerline{\includegraphics[width=0.75\linewidth]{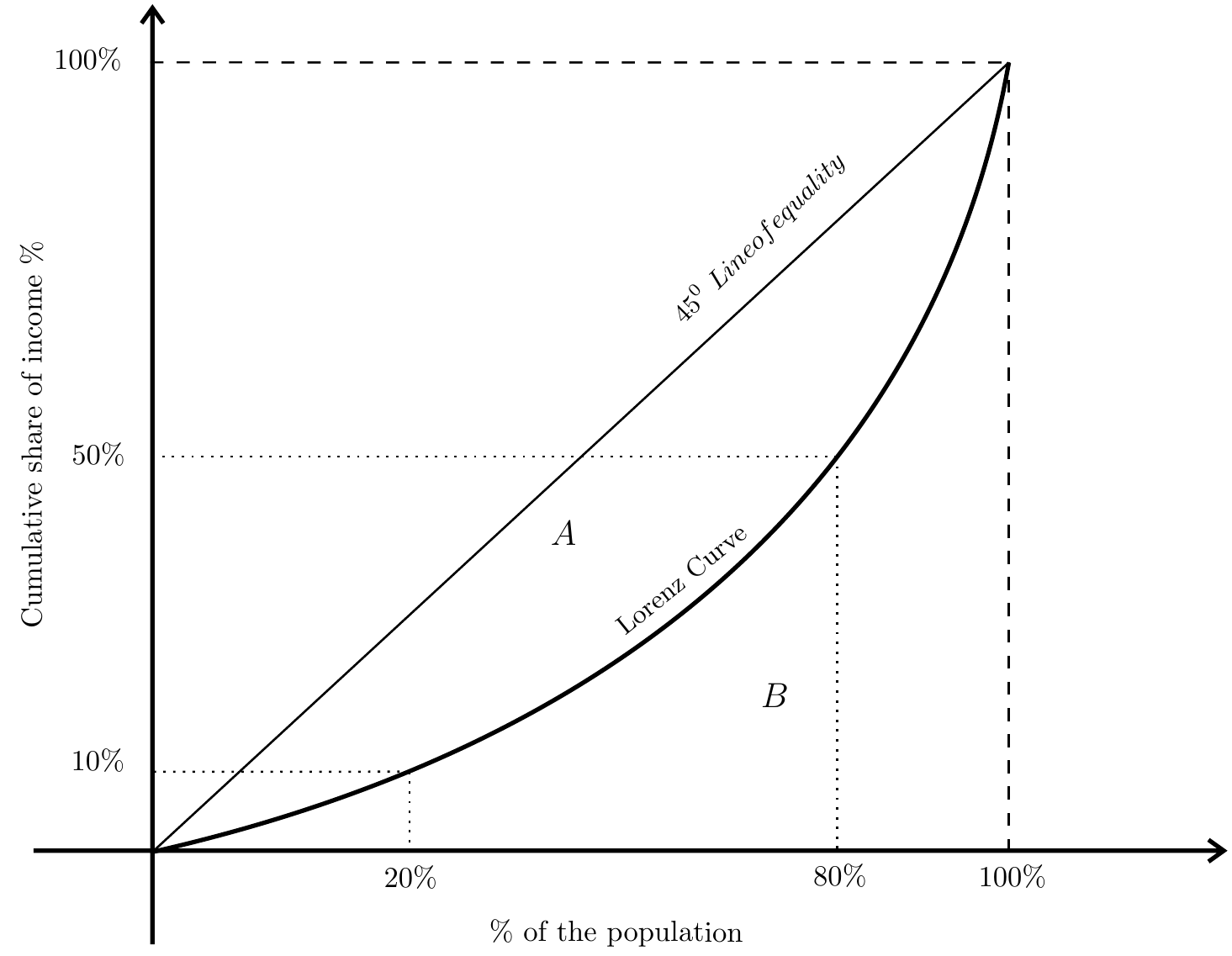}}
    \caption{\label{lcurve} The Lorenz curve framework (hypothetical data). }
\end{figure}

The Lorenz curve shows the percentage of total income earned by the cumulative percentage of the population. In a perfect income equality, the $25\%$ lowest income of the population would earn $25\%$ of the total income, the $50\%$ lowest income of the population would earn $50\%$ of the total income, hence the Lorenz curve would follow a 45° line. As inequality increases, the Lorenz curve deviates from the line of equality as shown in Fig. \ref{lcurve}.

The Gini coefficient is defined by the area between the equality line and the Lorenz curve divided by the total area below the equality line, that is, $G = A/(A + B)$. It is also equal to $1-2B$ due to the fact that $A + B = 0.5$.

For a continuous income probability the Lorenz curve $L(F)$ can be represented as a parametric function in $L(m)$ and $F(m)$, where $F$ is the cumulative distribution and $m$ is the income. The value of the area $B$ can be found by integration:

\begin{equation}\label{eq:B}
    B=\int\limits^{1}_{0} L(m)  dF(m)=\int\limits^{\infty }_{0} L(m)  P(m) dm ,
\end{equation}
where $P(m)$ is the probability density function, $\mu$ is the average income and

\begin{equation}
    L(m)=\frac{1}{\mu}\int\limits^{m}_{0} xP(x)  dx  ,
\end{equation}
being the percentage of total income by the population with up to income $m$. Simplifying Eq. \ref{eq:B} using integration by parts, the Gini coefficient becomes

\begin{equation}\label{eq:G}
   G(C)=1-\frac{1}{\mu}\int\limits^{\infty }_{0} C(m)^2 dm ,
\end{equation}
where $\mu$ is the average income of the $P(m)$ distribution and $C(m)=1-F(m)$ is the complementary cumulative distribution.

Using the formula above, an exponential distribution has a Gini coefficient of $0.5$. Therefore, for the two-class model, the Gini coefficient is a good indicator of how much the Pareto correction affects inequality. Other important property of Gini coefficient in this context is that it can be written as

\begin{equation}\label{gini-t}
    G( \lambda ,\alpha ) = 1 -\frac{\left( 1 - \lambda ^{2}\right) /2 - (\lambda ^{2}  \log  \lambda )/( 2\alpha  - 1)}{( 1 - \lambda )  - \lambda \log \lambda /( \alpha -1)},
\end{equation}
and expanding previous equation around $\lambda=0$, we arrive at
\begin{equation}\label{gini-e}
 G( \lambda ,\alpha ) =\frac{1}{2}  +\left[\frac{  \log\left( \lambda ^{-1}\right)}{( \alpha  -1)} -1\right]  \frac{\lambda }{2} +\mathcal{O}\left( \lambda ^{2}\log \lambda \right) .
\end{equation}
Hence, the Gini coefficient only depends on the Pareto index and the percentage of people that belong to the Pareto distribution.

Our model parameters and this theoretical Gini coefficient will be the set of indicators for analysis of inequality.

\section{Optimization of the two-class model with hybrid-PSO}

In this section we are going to detail the procedures to perform the Particle Swarm Optimization to fit the empirical CCDF by the two-class model. To better differentiate the theoretical and empirical variables or statistic, will be used the following notation

\begin{gather*}
x \longrightarrow  \textit{for empirical variables or statistic} \\
\hat{x} \longrightarrow  \textit{for theoretical variables or statistic}    
\end{gather*}

Firstly, we are going to calculate the empirical $CCDF$. Taking a sample $m_1,m_2...,m_N$ of income drawn from a population, in this case Brazilian income, thus $m_{(1)} \geqslant m_{(2)} \geqslant ... m_{(N)}$ is the income order statistic. Accordingly, the empirical CCDF of the PNAD income data is defined as

\begin{gather}\label{CD}
C(m_{n})= \frac{1}{N}\sum_{i=1}^{N} \mathbb{1}_{m_{(i)}\geqslant m_{n}}\\
\mathbb{1}_{m_{(i)}\geqslant m_{n}}=\begin{cases}
    &1 \quad m_{(i)}\geqslant m_{n}  \\
    &0 \quad m_{(i)}<m_{n} 
  \end{cases} , 
\end{gather}
where $N$ is the total number of people in the data and $I((m>m_i))$ is the indicator function.

After finding the cumulative distribution of the data we need to define our loss function \\ (quality measure). In the literature this is done separately for exponential and Pareto regions, fitted by minimizing a goodness-of-fit function after passing through a logarithmic transform \cite{2US,2BR}. Before specifying our loss function we will define two regularization terms to ensure meaningful values of $T$ and $\lambda$. So taking the theoretical income average for the exponential regime

\begin{equation}\label{ExpAvg}
    \langle\hat{m}\rangle _{exp} = T - m_c \Big[\exp\Big(\frac{m_c}{T}\Big)-1\Big]^{-1} 
\end{equation}
then the regularization being added to the loss function to ensure the equality between the theoretical and empirical average for the exponential region is of the form
    
\begin{equation}\\ \label{l1}
    l_1\stackrel{\text{def}}{=} \left|  \frac{T - m_c \Big[\exp\big(\frac{m_c}{T}\big)-1\Big]^{-1} }{\frac{1}{N_e}\sum ^{N_e}_{i=1}m_{(i)}}  -1\right| , 
\end{equation}
where $N_e$ is the greatest rank statistic (index of the ordered income) belonging to the exponential region. Therefore, the parameter $T$ that comes out of the process of optimization with this regularization can be interpreted as the estimation of the average income of the data, in the hypothetical case of an exponential distribution without any Pareto tail.

In parallel, note that can be a difference between the $\lambda = \hat{C}(m_c)$ and $C(m_c)$, which are the model top percentage and the data percentage of people gaining more than $m_c$, respectively. This is addressed by the second regularization term

\begin{equation}\label{l2}
    l_2\stackrel{\text{def}}{=} \left|\frac{\lambda}{C(m_c)}-1 \right|.
\end{equation}
Thus, the parameter $\lambda$ will always be equivalent to the percentage of people that have the power-law behavior given by the data.

Let $\eta_{n}\geqslant\eta_{n-1}...\geqslant\eta_{1}$ be an order statistic that follows $\eta_{n}=m_{\big(\lfloor n \frac{N}{k}\rfloor \big)}$\footnotemark where $n=1,2, ... ,k$ and $k<N$. This will be called class statistic and it divides the data into $k$ income points.

\footnotetext{The notation $\lfloor . \rfloor$ denotes the floor function, which can be formally define as $\lfloor x\rfloor=\max\{m \in \mathbb{Z} :m\leqslant x\}$}

Now we can define the measure of quality as a root mean squared logarithmic error (RMSLE) between the data and the model using the $\eta_n$ statistics

\begin{gather}\label{RMSLE}
    \text{RMSLE}(C(\eta_{n}),\hat{C}(\eta_{n},\pmb{x}))= \sqrt{\frac{1}{k-1}\sum ^{k-1}_{n=1}\Bigl\{ \log\bigl[C(\eta_{n})\bigr] - \log\bigl[ \hat{C}(\eta_{n},\pmb{x})\bigr]\Bigr\}^2}, 
\end{gather}
where $C(\eta_{n})$ is the empirical complementary cumulative distribution, $\hat{C}(\eta_{n},\pmb{x})$ is the CCDF of the model and $\pmb{x}$ is the parameter vector. Using the class statistic not only helps with the computational load, but also gives consistency to  the CCDF precision used in the loss function. With $m_{(n)}$ statistics the precision of the CCDF was tied to the number of the population, having $N$ points. The set $\eta_n$ instead has $k-1$ points, which is independent on the population number. Note that the $\eta_k$ was excluded from the loss function since $\log\bigl[C(\eta_{k})\bigr]$ diverges. In this paper $k=10000$.

Therefore using equations (\ref{l1}), (\ref{l2}) and (\ref{RMSLE}) we define the loss function

\begin{align}\label{loss}
\text{Loss}(\pmb{x})=\text{RMSLE}(C(\eta_{n}),\hat{C}(\eta_{n},\pmb{x}))  + l_1 + l_2 .
\end{align}

Now, a defined search space is needed, in order for the PSO to find a solution. To properly determine the crossover region, we are going to use the empirical crossover percentage $p=C(m_c)$, that is the empirical CCDF evaluated in the income $m_c$. Note that the income variable is not suited to separate the data since it has a lot of repeated values. With this percentage we calculate the crossover income $m_c$ from a linear interpolation of the empirical CCDF (Eq. \ref{CD}). The interpolation transforms the empirical cumulative distribution into a continuous distribution, thus a solution $m_c=C^{-1}(p)$ will not have discrete values. Therefore given a $T$ with Eq. \ref{sigmaM} we can determine $\lambda$, then for a vector of parameters $\pmb{x}=(C(m_c),\alpha,T)$ we can define a theoretical CCDF ($\hat{C}(m)$) using Eq. \ref{CM}, allowing us to derive a goodness-of-fit function.

Another information that needs to be provided, in order to define the search space is its range. For our case $T\in\big[\frac{\langle m\rangle}{2},2\langle m\rangle\big] $, $\alpha \in [1,3]$ and $C(m_c) \in  (0, 0.2]$. Therefore the search space is a cuboid $T$ in the parameter coordinates. The next step is to define the optimizer that will minimize the loss function (\ref{loss}).

The Particle Swarm Optimization (PSO) is a computational method that optimizes a problem by iteratively trying to improve a set of candidates in the parameter space (in our case $\{C(m_c),\alpha,T\}$) with regard to a given measure of quality \cite{Pso}. Let $\{\tensor{\pmb{x}}{^i_{t}}=(\tensor{C(m_c)}{^i_{t}},\tensor{\alpha}{^i_{t}},\tensor{T}{^i_{t}})\}$ with $i=1,2,...N_c$, where $N_c$ is the number of candidates of parameters, be our set of candidate vectors in the $t_{th}$ PSO step. Each candidate is treated as a solution of the problem. Thus the optimization will be derived by the search of the parameter space with $N_c$ candidates. The best solution will minimize the quality measure. In the first iteration each candidate index is part of $K$ randomly chosen sets $S_0 \in \{S^i_0\}_{i\in\{1,2...,N_c}\}$, this process defines the set of neighbors of all candidates. $S^i_{t-1}$ is the neighbors set of the $i_{th}$ candidate in the $t_{th}$ step and it contains the neighbors index and its own index ($i$). One candidate index can randomly choose to participate in a set repeatedly times and duplicated indexes are removed, thus $S^i$ size may vary. These neighbors sets are redefined by the same random process in each step that the algorithm didn't improve the best solution between the history of all the candidates. 

This sets $S^i_t$ containing randomly chosen candidates indexes are used to inform a specific property of those to the $i_{th}$ candidate. The exploration will use these sets to compose the next step of each candidate and it will be clarified in the next subsection. The 2007 standard PSO (SPSO2007) value of $K=3$ and will be used for this paper.

The exploration is done by each candidate making steps that are influenced by the direction to the best neighbors position (the best position between its set of neighbors and itself), the candidate best position in its step history and its last step direction. For each candidate $i$, the step $t$ is determined by:

\begin{gather}
\tensor{\pmb{v}}{^i_{t}}=W \tensor{\pmb{v}}{^i_{t-1}}+c_1 \Lambda_1(\tensor{\pmb{P}}{^i_{t-1}}-\tensor{\pmb{x}}{^i_{t-1}})+c_2 \Lambda_2(\tensor{\pmb{G}}{^i_{t-1}}-\tensor{\pmb{x}}{^i_{t-1}})\\
\tensor{\pmb{x}}{^i_{t}}=\tensor{\pmb{x}}{^i_{t-1}}+\tensor{\pmb{v}}{^i_{t}} ,
\end{gather}
where $\pmb{x}$ is the vector position, $\pmb{v}$ is analog to the velocity,
\begin{equation}
\tensor{\pmb{P}}{^i_{t}}=\{ \pmb{x} \in \{\tensor{\pmb{x}}{^i_{\nu}}\} : \text{Loss}(\pmb{x})=\min_{\pmb{z} \in \{\tensor{\pmb{x}}{^i_{\nu}}\}} \text{Loss}(\pmb{z}) \}     
    \end{equation}
with $\nu=0,1,...,t$ is the personal best in regard to the quality measure,
\begin{equation}
\tensor{\pmb{G}}{^i_{t}}=\big\{ \pmb{x} \in C^i_t : \text{Loss}(\pmb{x})=\min_{\pmb{z} \in C^i_t} \text{Loss}(\pmb{z}) \big\}
\end{equation}
where $C^i_t\equiv\{\tensor{\pmb{P}}{^j_{t}}\}_{j \in  \{S^i_t\}}$ is the set of personal best of neighbors set of the $i_{th}$ candidate in the $t_{th}$ step and $\tensor{\pmb{G}}{^i_{t}}$ is the best position in the $C^i_t$ set.
The parameters $\Lambda_1$ and $\Lambda_2$ are uniformly random with range $[0,1]$. The $W$, $c_1$ and $c_2$ are considered hyperparameters, so they will have a fixed value (or a behavior predetermined). In this paper $c_1=c_2=1.7$ and $W$ fall linearly in  $[0.7,0.4]$. The initialization is done as follows:

\begin{equation}
    \begin{gathered}
        \begin{cases}
    \tensor{\pmb{x}}{^i_{0}}=\pmb{U}_T\\
    \tensor{\pmb{v}}{^i_{0}}=\displaystyle{\frac{\pmb{U}_T-\tensor{\pmb{x}}{^i_{0}}}{2}}\\
    \tensor{\pmb{P}}{^i_{0}}=\tensor{\pmb{x}}{^i_{0}}\\
    \tensor{\pmb{G}}{^i_{0}}=\big\{\pmb{x} \in \{\tensor{\pmb{P}}{^j_{0}}\}_{j \in  \{S^i_0\}} : \text{Loss}(\pmb{x})=\min_{\pmb{z} \in \{\tensor{\pmb{P}}{^j_{0}}\}} \text{Loss}(\pmb{z}) \big\}
        \end{cases},
    \end{gathered} 
\end{equation}
where $\pmb{U}_T$ is a random vector inside the search space cuboid drawn according to the uniform distribution.

PSO was chosen for being able to work with non-differentiable error function and discrete variables \cite{IgorNon}, which is needed since the two-class loss function is not differentiable. The standard PSO does not always converge to a good solution, so a hybrid approach was used \cite{hybrid}. This hybrid approach utilizes L-BFGS-B steps to improve the candidates local search ability (exploitation). Even though BFGS is a Quasi-Newton method it can be used for nonsmooth optimization \cite{BFGSnon1,BFGSnon2}.

In short, to find the model parameters of section 2, first we need to calculate the cumulative probability distribution of income for the Brazilian population (\ref{CD}) and then use hybrid-PSO to fit the model by minimizing the value of the function (\ref{loss}). The results from these procedures will be displayed in the next section.

\section{Results and cross-validation}    

In this section, we are going to perform cross validation with the bootstrap method in order to test the accuracy of the model \cite{BootCV}. As mentioned above, the bootstrap sampling is performed before calculating the cumulative distribution, that is, picking a random sample of the income data, with replacement, then calculating the cumulative distribution. This approach not only will give us the ability to estimate model parameters errors as well as will allow us to do an out of the bootstrap cross-validation (BCV) \cite{outboot}. On average, random sampling with replacement includes $(1-e^{-1})\%\approx 63.2\%$ of the original data in each bootstrap set of samples and hence the rest allow us to define out-of-sample test sets.

\begin{table}[H]
   \caption{\label{tb:Rtable} Mean, standard deviation, $95\%$ confidence interval and coefficient of variation of the model parameters, Gini coefficient, and RMSLE of the training and test sets. 
%  Every pair of training set and test set is defined with bootstrap samples, being the training set the selected sample and the test set the rest of the original set. The bootstrap sampling was repeated 1000 times.
} 
\begin{center}

    \begin{tabular}{llllllcll}
    \toprule 
     \multicolumn{7}{l}{$\displaystyle 2019$ Data} & \multicolumn{1}{c}{} & \multicolumn{1}{c}{} \\
    \midrule 
     \multicolumn{1}{c}{} & \multicolumn{1}{c}{} & $\displaystyle mean$ &  & $\displaystyle \sigma $ &  & {\footnotesize $\displaystyle 95\%\ CI$} &  & \multicolumn{1}{c}{$\displaystyle CV$} \\
    \hline 
     Crossover Income (R\$) &  & $\displaystyle 3977$ &  & $\displaystyle 42$ &  & $\displaystyle 3900\ //\ 4000$ &  & $\displaystyle 1.06\times 10^{-2}$ \\
    Top-percentage (\%) &  & $\displaystyle 10.64$ &  & $\displaystyle 0.26$ &  & $\displaystyle 10.42\ //\ 11.17$ &  & $\displaystyle 2.45\times 10^{-2}$ \\
    Temperature (R\$) &  & $\displaystyle 1775$ &  & $\displaystyle 3.40$ &  & $\displaystyle 1769\ //\ 1782$ &  & $\displaystyle 1.92\times 10^{-3}$ \\
    Pareto index &  & $\displaystyle 1.789$ &  & $\displaystyle 0.011$ &  & $\displaystyle 1.767\ //\ 1.782$ &  & $\displaystyle 6.04\times 10^{-3}$ \\
    Gini coefficient &  & $\displaystyle 0.578$ &  & $\displaystyle 0.001$ &  & $\displaystyle 0.576\ //\ 0.581$ &  & $\displaystyle 2.31\times 10^{-3}$ \\
    \hline\hline 
     Train set RMLSE &  & $\displaystyle 0.1486$ &  & $\displaystyle 0.0008$ &  & $\displaystyle 0.1470\ //\ 0.1501$ &  & $\displaystyle 5.31\times 10^{-3}$ \\
    Test set RMLSE &  & $\displaystyle 0.1489$ &  & $\displaystyle 0.0017$ &  & $\displaystyle 0.1458\ //\ 0.1525$ &  & $\displaystyle 1.12\times 10^{-2}$ \\
     \hline
    \end{tabular}
    \end{center}

\end{table}

In this context, firstly, we define a pair of sets by bootstrap sampling: training set and test set. Then, with the CCDF of the training set, we use the hybrid-PSO to find the optimal parameters. Lastly, using the optimal parameters evaluated above, we calculate the RMSLE of each set. Lastly, we calculate the RMSLE of each set using the optimal parameters evaluated above. Repeat this process for $R$ pairs of training and test sets, in this study, we have chosen $R = 1000$. This procedure gives us the ability to see how well our model can fit the CCDF of an income data set, which came from the same distribution of the training set but was not in the training process. 

In this section, we used from the data of the 2019 Continuous National Household Sample Survey (PNADc) available by the Brazilian Institute of Geography and Statistics (IBGE) since 2012. The PNADc is a research that collects data from a multitude of Brazilian social characteristics including labor, income and education. From PNADc data, we extracted the total monthly income \footnotemark, gender and color/race of each person in the year 2019.  We neglected people without income and missing values. 

\footnotetext{PNADc microdata variables are organized with codes, the prefix V is for a pure variable that is extracted directly from the survey and the VD is for the composed variable, usually a linear equation of pure variables. For the total monthly income we use the variable VD$4022$, which is the total income from all sources, including aid programs among others. This variable is only present in the first annual interview of PNADc.}

%As said before, the model was fitted with a varying $a$ with 40 bootstrap sets and obtain $a=1.0084\pm0.0403$, corroborating with our choice of fixing $a=1$, making the model more robust.

After training the parameters of our model, in order to complete our set of indicators, we calculate the theoretical Gini coefficient following section \ref{secgini}.

As shown in Table \ref{tb:Rtable} the mean RMSLE of the test set is close to the training set RMSLE, so the model has little bias. The parameters of the model have small coefficients of variation (CV), and the highest is that of the Top-percentage. This is due to the discontinuity of the empirical income data at the crossover.

\begin{figure}[H]
    \centering
    \centerline{\includegraphics[width=0.85\linewidth]{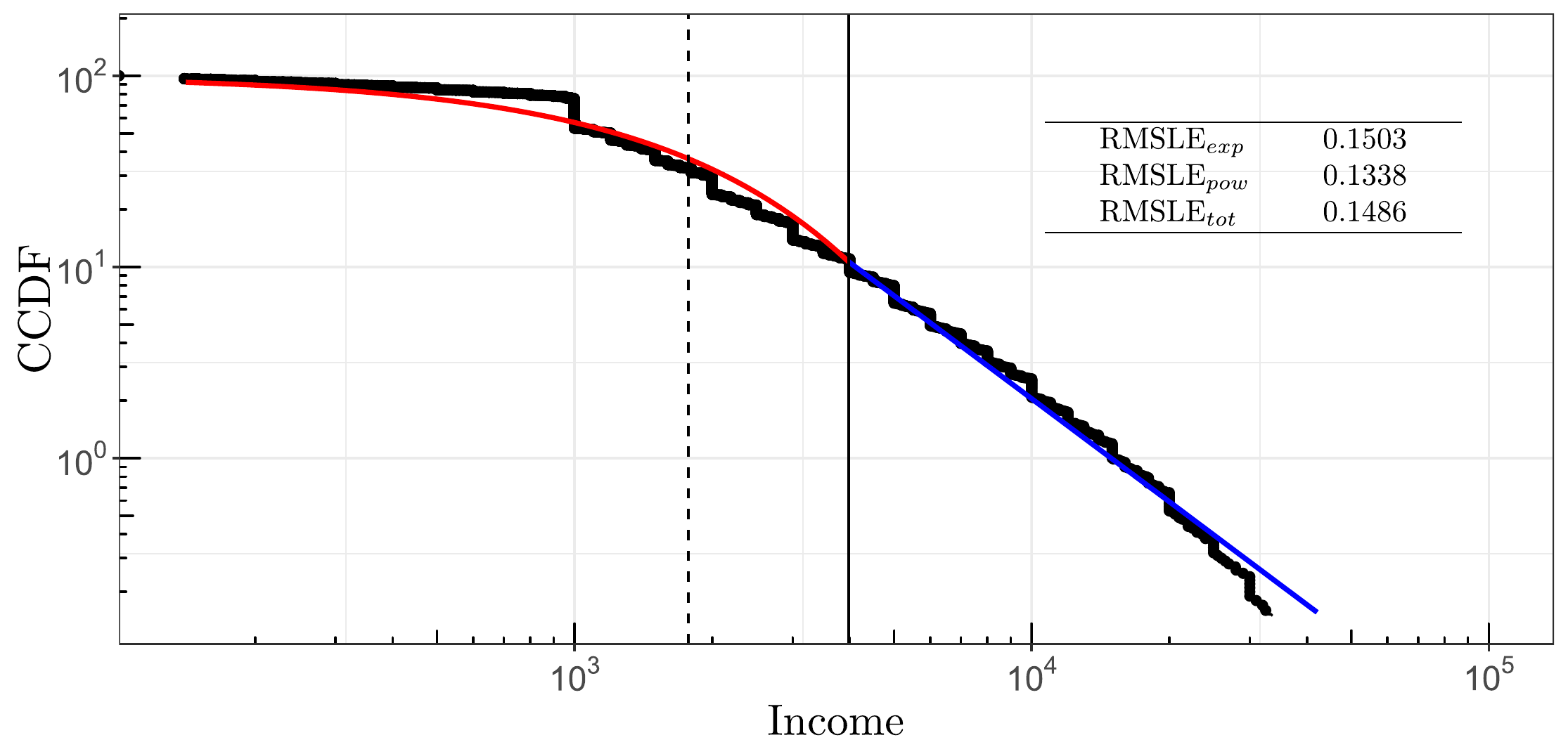}}
    \caption{\label{fig:fit}Cumulative probability distribution of income on a log–log scale. The black points represent the cumulative distribution of the data from 2019 PNAD and the solid lines correspond to the fitted model described in Eq. \ref{CM}. The red curve obeys Boltzmann–Gibbs distribution and the power-law distribution is characterized by the blue curve. The temperature is highlighted by the dashed vertical line, at value $\text{R\$ }1775 \pm 3$, and the crossover income is represented by the solid vertical line at $m_c = \text{R\$ } 3977\pm 42$, which separates the Boltzmann–Gibbs and Pareto regions of income. Therefore $\lambda= (10.64\pm 0.26)\%$ is the top income percentage of people which obeys a Pareto power-law with index  $\alpha = 1.789 \pm 0.011$. Also, the RMSLE of the original data set calculated in each part of the distribution and total can be found in the top right table. }
\end{figure}

Fig. \ref{fig:fit} displays the cumulative distribution fitted with our model with the parameters shown in Table \ref{tb:Rtable}. These results were estimated by bootstrapping the data, calculating the cumulative distribution for each bootstrap set and then fitting the model numerically with PSO. Analyzing the RMSLE between the model with estimates of the parameters and the original data set in parts, we can identify that the Pareto region have a greater error, which is expected since Pareto part does not capture the top $0.01\%$ very well, and since the error (Eq. \ref{loss}) is logarithmic, outliers whose $C(m)$ values are close to $0$ are amplified. In the exponential part, a fraction of the error is due to the minimum wage effect (low income discontinuity at $m=\text{R\$ }998$) as can be seen in Fig. \ref{fig:fit}. 

Comparing this method with a crossover income being determined with a fixed proportion for the Pareto tail, $\lambda=5\%$ , as seen in reference \cite{2BR}, we get a temperature equals to the mean, $T=2067\pm 7$, a Pareto index of $1.816\pm 0.006$  and a $m_c=6192\pm 21$. This approach displays discontinuity between BG and Pareto regions, its training set error is $0.2257\pm 0.002$ and the test set error is $0.2258\pm 0.002$, which are significantly higher than with an optimal crossover (Table \ref{tb:Rtable}).

We applied our model to the data of 2019 stratified into two dichotomies of the population allowing us to compare each indicator and test our model in different data distributions. The first dichotomy is the division by gender(man and woman) and the second is the division by race/color  (black/brown/indigenous - BBI - and white/yellow - WY)\footnotemark. The results and validation can be seen in the Table \ref{tb:SVtable}.

\footnotetext{Brown stands for mixed race and yellow stands for asian.}

\begin{table}[H]
    \caption{\label{tb:SVtable} Inequality indicators and RMSLE for a stratified data.} 
\begin{center}

\begin{tabular}{lllllll}
\toprule 
 \multicolumn{7}{l}{Stratified $\displaystyle 2019$ Data } \\
\hline 
 Categories &  & \multicolumn{2}{l}{Sex/Gender} &  & \multicolumn{2}{l}{Race/Color} \\
\cmidrule{3-4} \cmidrule{6-7} 
 Subgroup &  & Man & Woman &  & WY & BBI \\
\cmidrule{3-4} \cmidrule{6-7} 
 Crossover Income (R\$) &  & $\displaystyle 4930\pm 13$ & $\displaystyle 3503\pm 25$ &  & $\displaystyle 4987\pm 55$ & $\displaystyle 3995\pm 26$ \\
Top-percentage (\%) &  & $\displaystyle 8.97\pm 0.59$ & $\displaystyle 9.93\pm 0.16$ &  & $\displaystyle 11.76\pm 0.30$ & $\displaystyle 6.42\pm 0.12$ \\
Temperature (R\$) &  & $\displaystyle 2047\pm 6$ & $\displaystyle 1517\pm 4$ &  & $\displaystyle 2329\pm 7$ & $\displaystyle 1455\pm 3$ \\
Pareto index &  & $\displaystyle 1.72\pm 0.02$ & $\displaystyle 1.89\pm 0.02$ &  & $\displaystyle 1.74\pm 0.01$ & $\displaystyle 2.01\pm 0.02$ \\
Gini coefficient &  & $\displaystyle 0.584\pm 0.002$ & $\displaystyle 0.565\pm 0.002$ &  & $\displaystyle 0.588\pm 0.002$ & $\displaystyle 0.547\pm 0.001$ \\
\hline\hline 
 Train set RMLSE &  & $\displaystyle 0.155\pm 0.001$ & $\displaystyle 0.155\pm 0.001$ &  & $\displaystyle 0.137\pm 0.001$ & $\displaystyle 0.177\pm 0.001$ \\
Test set RMLSE &  & $\displaystyle 0.156\pm 0.002$ & $\displaystyle 0.155\pm 0.002$ &  & $\displaystyle 0.138\pm 0.002$ & $\displaystyle 0.178\pm 0.002$ \\
 \hline
\end{tabular}
\end{center}  

\end{table}

Looking at the gender dichotomy, there is a significant difference between the theoretical Gini coefficient. Man’s income ”temperature” is considerably higher than the woman’s, but their Pareto and top-percentage indicators are lower, meaning that there is a less percentage of men in the Pareto Region, but their inequality in this region is higher. Remember that, for the two-class model, the temperature does not affect the theoretical Gini coefficient, so the properties of the Pareto region completely define the inequality.

Analyzing the color/race dichotomy, we get a higher contrast in their inequality indicators compared to the gender dichotomy. WY has top-percentage and Temperature considerably higher than BBI. BBI has the lowest top-percentage value, $\lambda = 6.42 \pm 0.12$, and the highest Pareto index, $\alpha = 2.01 \pm 0.02$, compared to all subsets. These results gives BBI the lowest Gini coefficient, indicating the highest equality within the subgroup.

\section{Temporal evolution of inequality indicators}

The time evolution of inequality indicators, in the context of the two-class model or the lognormal-Pareto model, is a subject of interest in the literature  \cite{2BR,USt,JPt}. However, the crossover income is usually fixed or determined by a log-log graph. In this paper, we introduce a formal approach to determine the temporal evolution of the optimal crossover income.

To be able to analyse the Brazilian inequality over the years, we applied our model to describe the income distribution between $2001-2019$. For our empirical data we used the National Household Sample Survey (PNAD) for the years 2001 to 2011, and for 2012 to 2019 we used the PNADc. The PNAD is the predecessor of the PNADc and was discontinued in 2015. We gave priority to PNADc in the years that the two survey programs were running since PNADc gives a broader territorial coverage and larger population sample.  We choose the first annual interview of PNADc since it contains the income of all sources. Following the same data cleaning procedure, we neglected people without income and missing values.

\begin{figure}[H]
    \centering

    \begin{subfigure}{0.6\linewidth}
        \includegraphics[width=\linewidth]{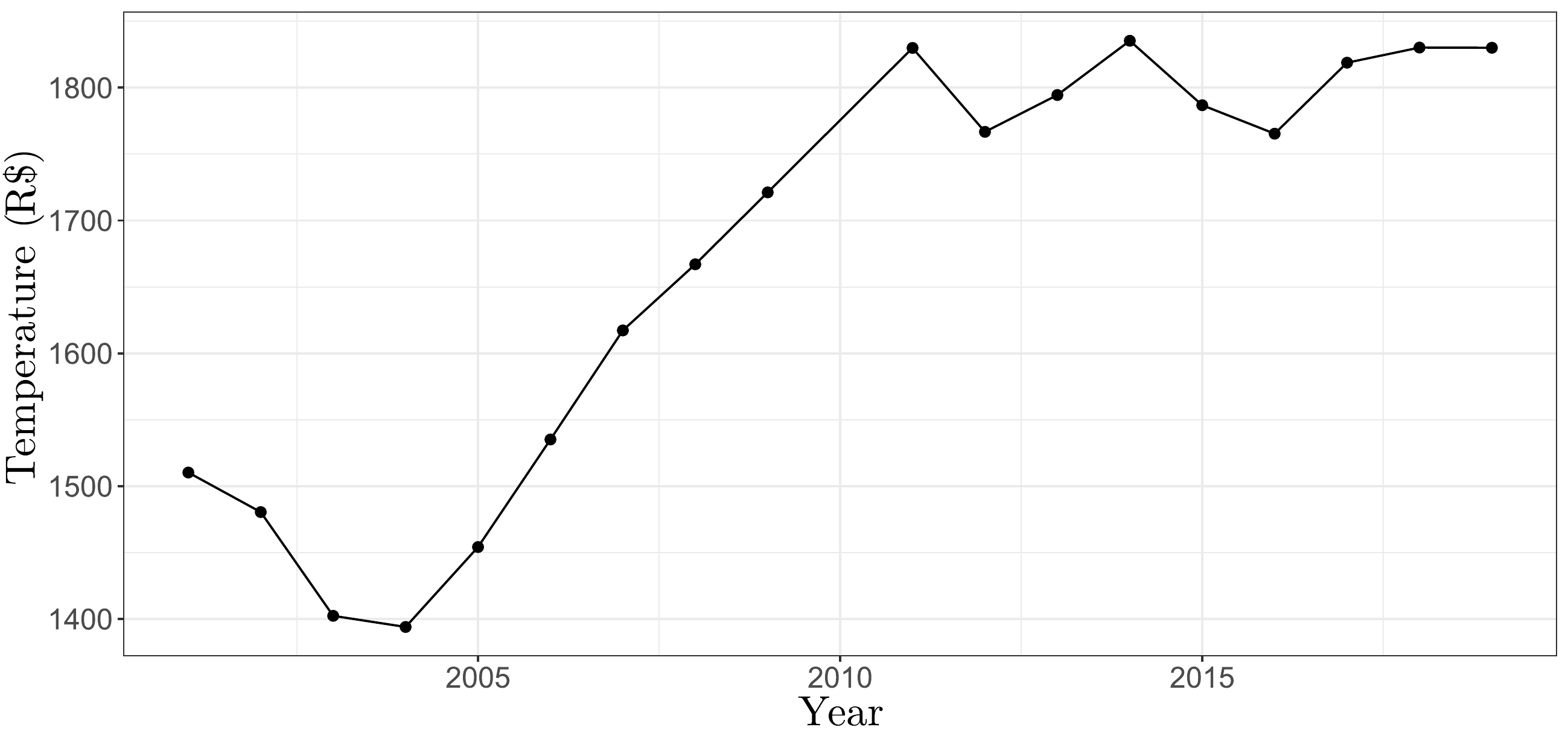}
        \caption{Deflated Temperature series.}\label{fig:T}
    \end{subfigure}
    \\ 

    \begin{subfigure}{0.6\linewidth}
        \includegraphics[width=\linewidth]{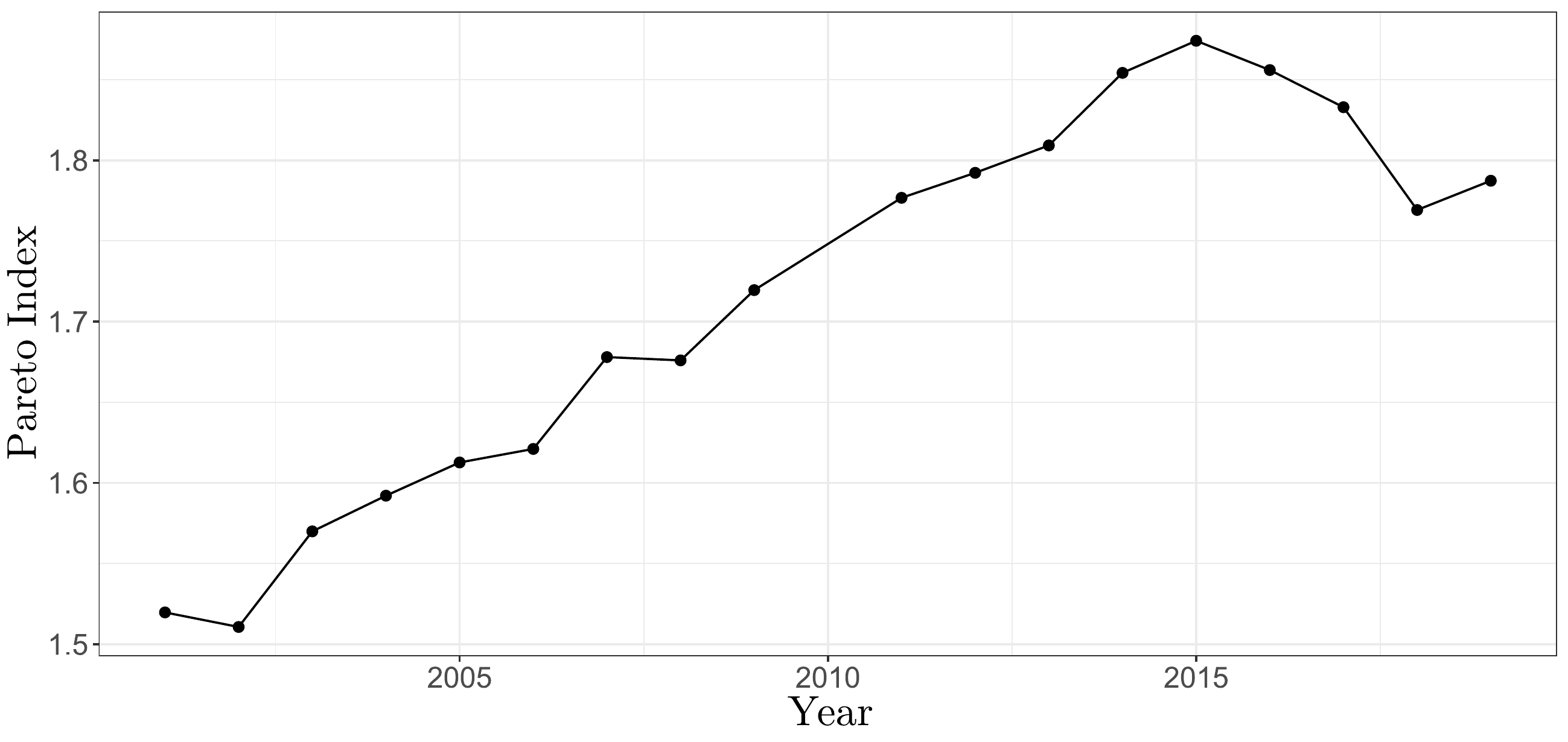}
        \caption{Pareto index series.}\label{fig:pareto}
    \end{subfigure}
    \\ 

    \begin{subfigure}{0.6\linewidth}
        \includegraphics[width=\linewidth]{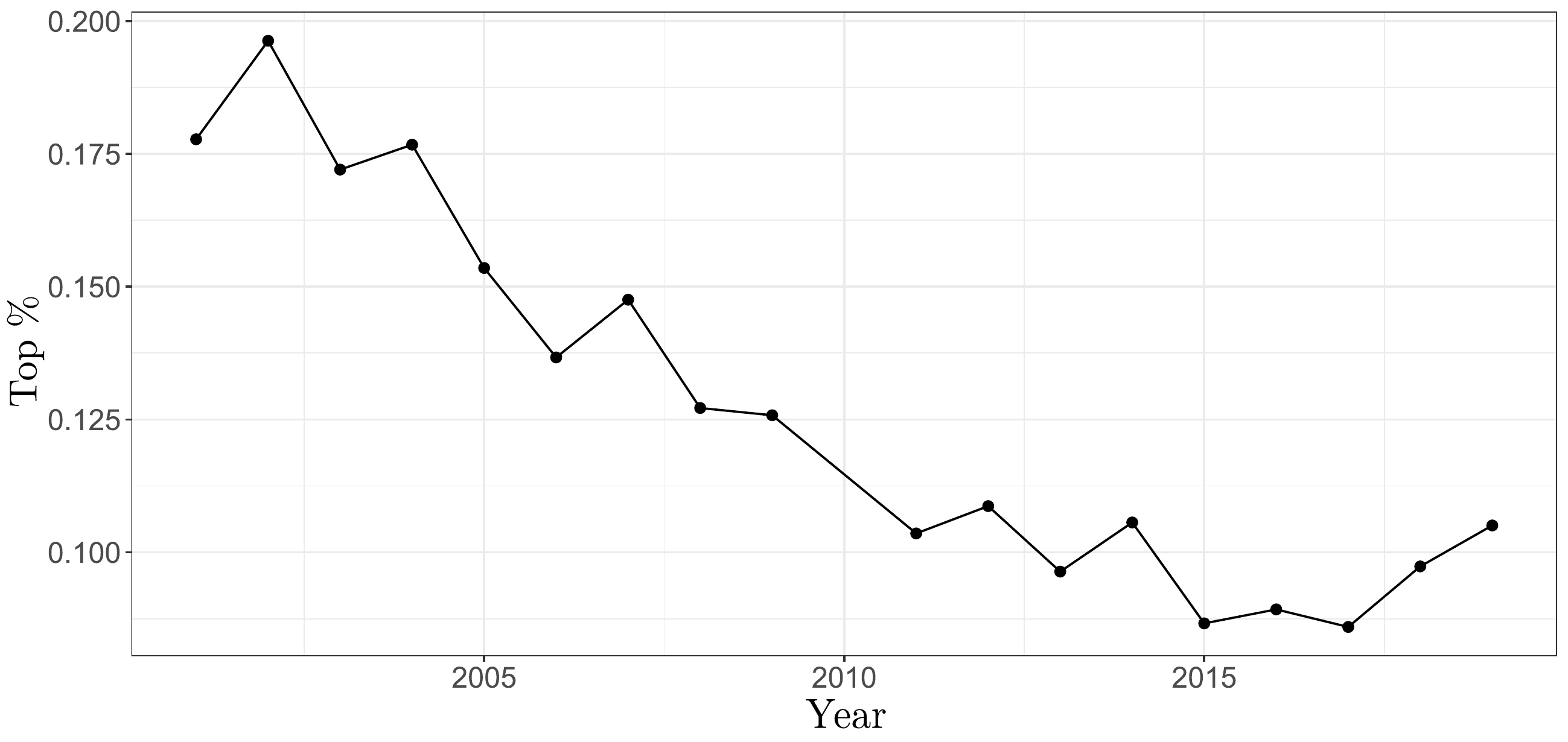}
        \caption{Top-percentage series.}\label{fig:top}
    \end{subfigure}
    \\

    \begin{subfigure}{0.6\linewidth}
        \includegraphics[width=\linewidth]{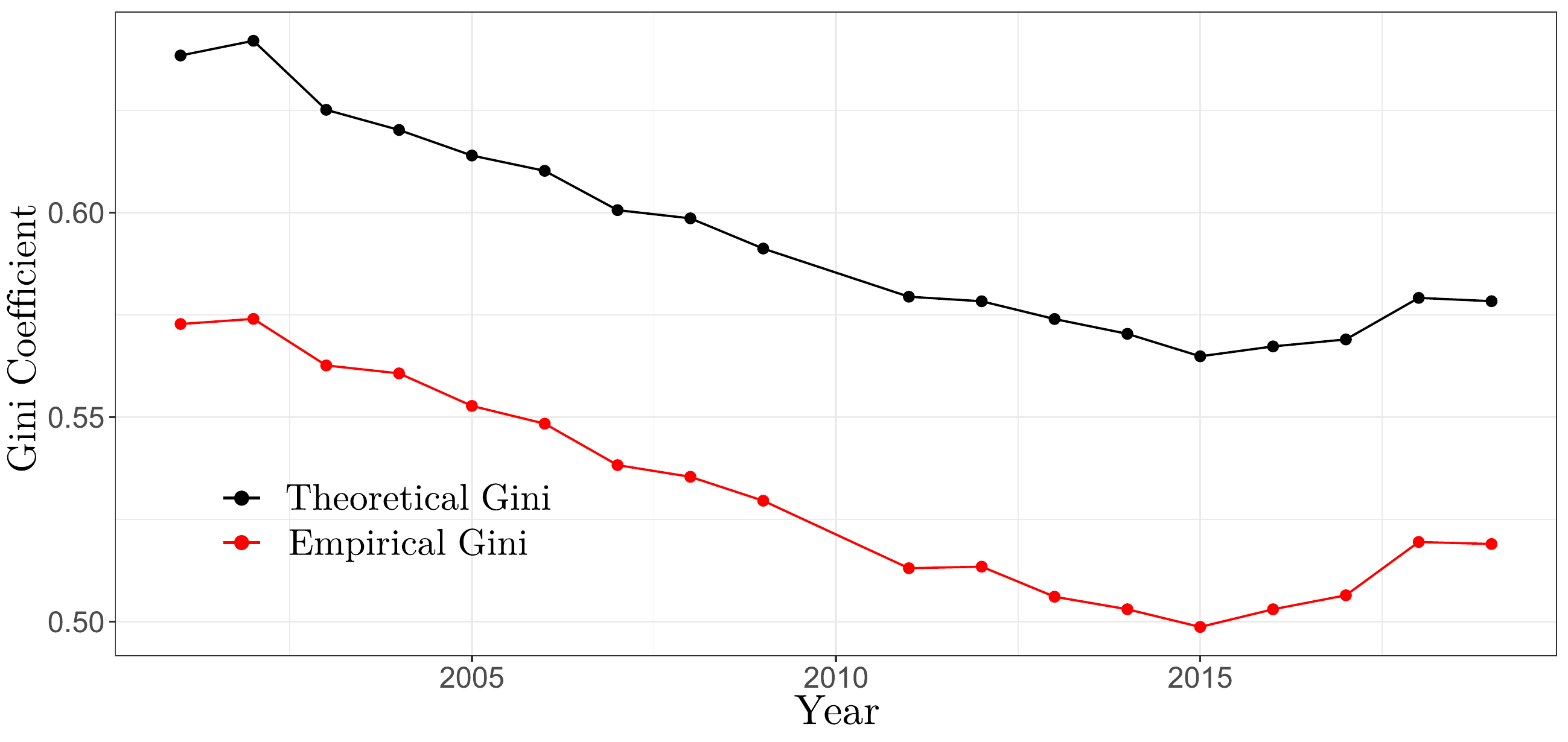}
        \caption{Gini coefficient series. The black points are the theoretical Gini, Eq. \ref{gini-t}, and the red points are the empirical Gini.}\label{fig:gini}
    \end{subfigure}
    \\ 

    \caption{Time series of the inequality indicators.}
    
\end{figure}

Lastly, we applied the optimization to fit the two-class model, Eq. \ref{CM}, to the data of each year. We can access the temporal evolution of each indicator in the Figs. \cref{fig:pareto,fig:top,fig:T,fig:gini}. We can see that every parameter of our model has an interpretable evolutionary trend. 

\vspace{10pt}
The temperature reflects the income power of the lower to middle class. This parameter was deflated to the $2019$ currency using the Broad Consumer Price Indices (IPCA), available by IBGE. According to the Fig. \ref{fig:T} there is trend of an increase in the temperature.  

\begin{figure}[H]
    \centering

    \begin{subfigure}{0.7\linewidth}
        \includegraphics[width=\linewidth]{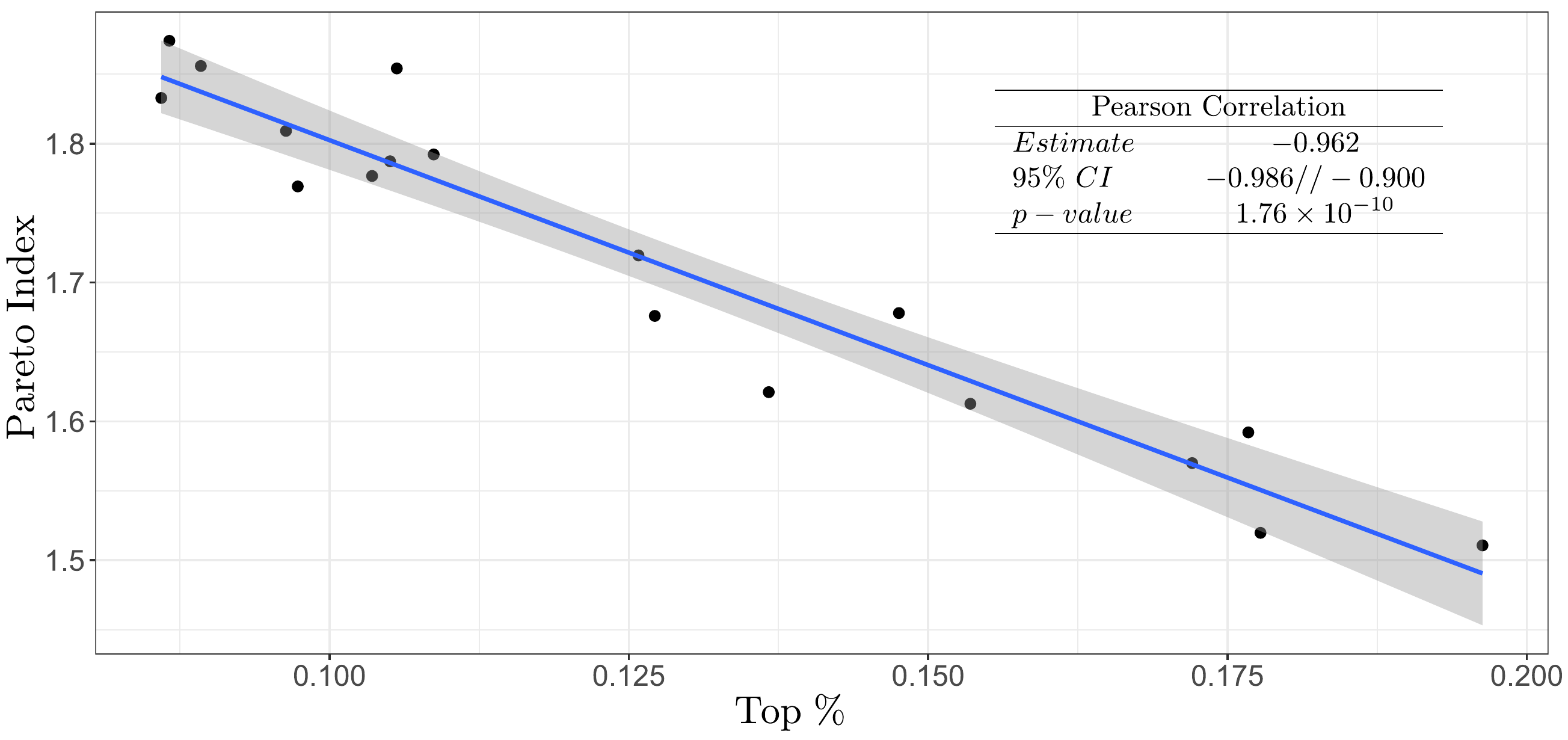}
        \caption{Correlation and affine regression with the Pareto index as a function of the top-percentage.}\label{fig:t-p}
    \end{subfigure}
    \\ \vspace{10pt}

    \begin{subfigure}{0.7\linewidth}
        \includegraphics[width=\linewidth]{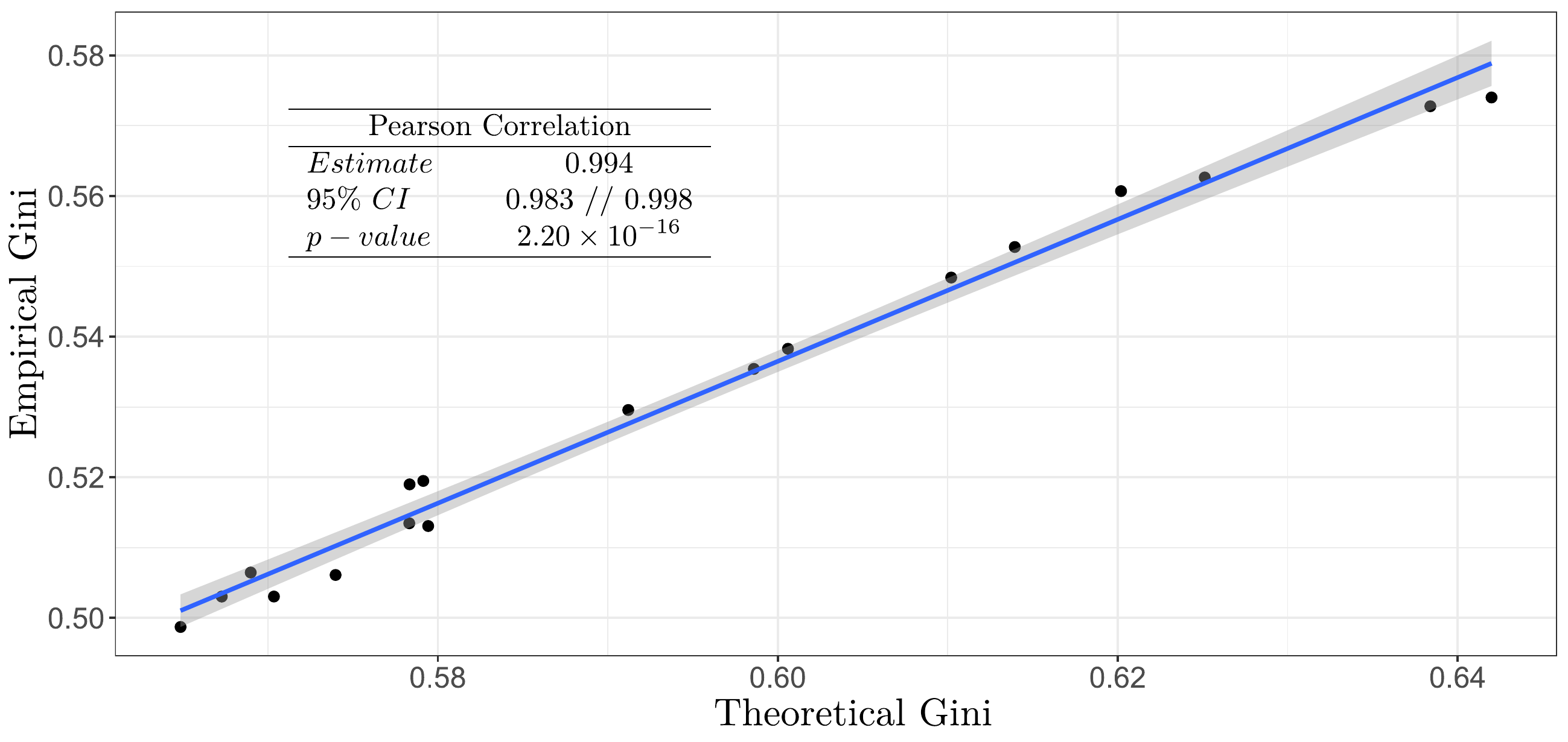}
        \caption{Correlation and affine regression with theoretical Gini as a function of the empirical Gini.}\label{fig:affine}
    \end{subfigure}
    \\

    \caption{Correlation and affine regression between indicators.}
    
\end{figure}

The Pareto index and the top-percentage have a strong anticorrelation as can be seen in the Fig. \ref{fig:t-p}. From $2002$ to $2015$, there is an increase in the Pareto index and a decrease in the top-percentage. This means that, according to our survey data, we are in the presence of an income redistribution process. And it is corroborated by the decrease of the Gini coefficient in this time range, see Fig \ref{fig:gini}.

As discussed above, the model does not capture the minimum wage effect on the distribution as well as the extreme high income, top $0.01\%$, where the Pareto behavior breaks down. These observations coupled with the fact that the empirical estimator suffers from a downward bias when the distribution used is fat-tailed \cite{taleb}, explains why the theoretical Gini coefficient is higher than the empirical one. Although Fig. \ref{fig:gini} shows an apparent big difference between theoretical and empirical Gini coefficients, it can be shown that they are very strongly correlated as can be seen in the Fig. \ref{fig:affine}. 

Fig. \ref{fig:affine} also shows the affine relation between the theoretical and the empirical Gini coefficients, as well as the Pearson correlation. The correlation is $\rho = 0.994$ with a p-value $2.20\times 10^{-16}$, which means that there is high evidence of the strong correlation between the two coefficients. The affine regression $G_{e} =( -0.07\pm 0.02) +( 1.01\pm 0.03) G_{t}$ between theoretical and empirical Gini has a residual standard error of $0.0029$. Since the theoretical and empirical Gini coefficients have a strong correlation, we can  conclude that the theoretical Gini can also be used as a measure of inequality, if the data can be well explained by the two-class model.

\begin{figure}[H]
    \centering

    \begin{subfigure}{0.5\linewidth}
        \includegraphics[width=\linewidth]{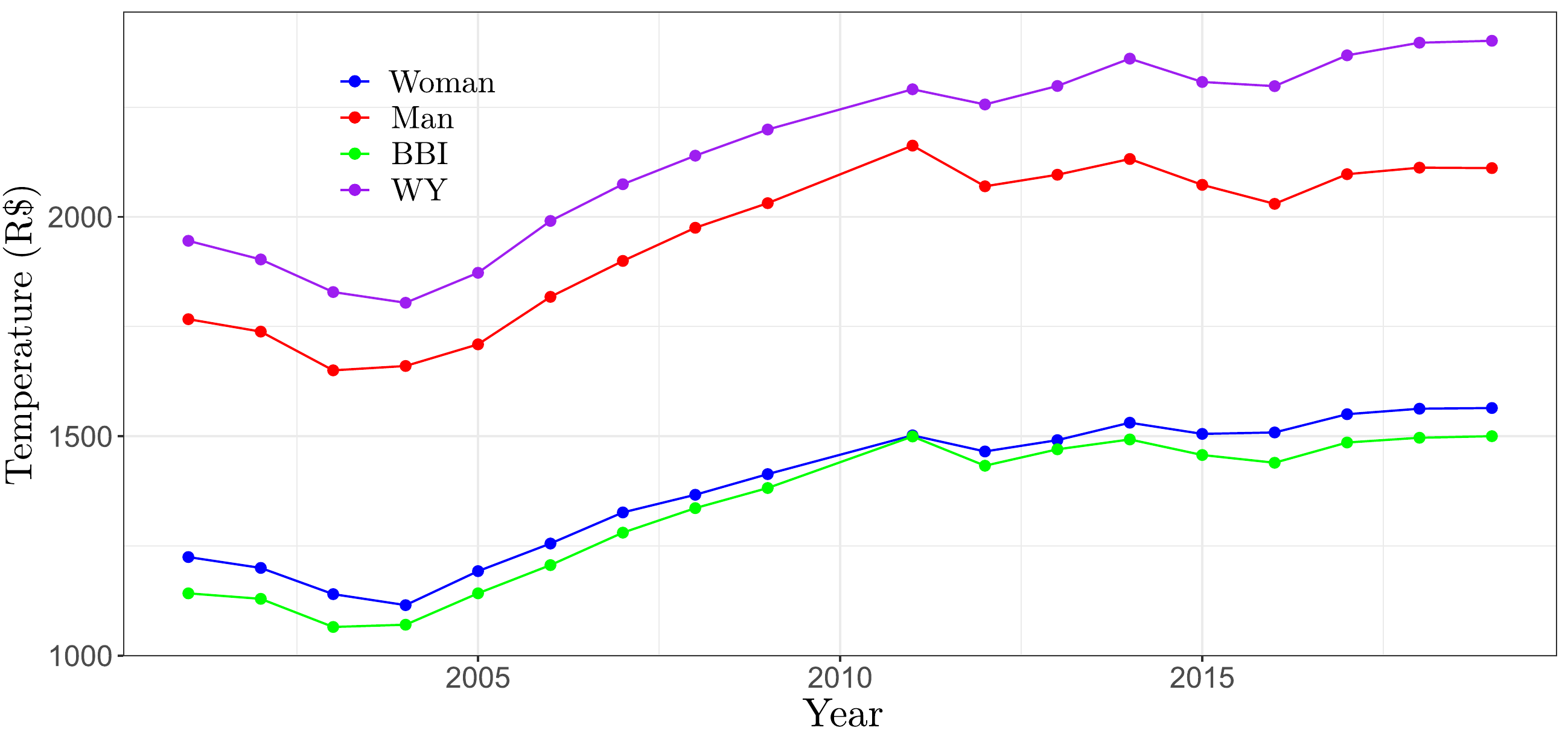}
        \caption{Deflated Temperature series.}\label{fig:ST}
    \end{subfigure}
     \\

    \begin{subfigure}{0.5\linewidth}
        \includegraphics[width=\linewidth]{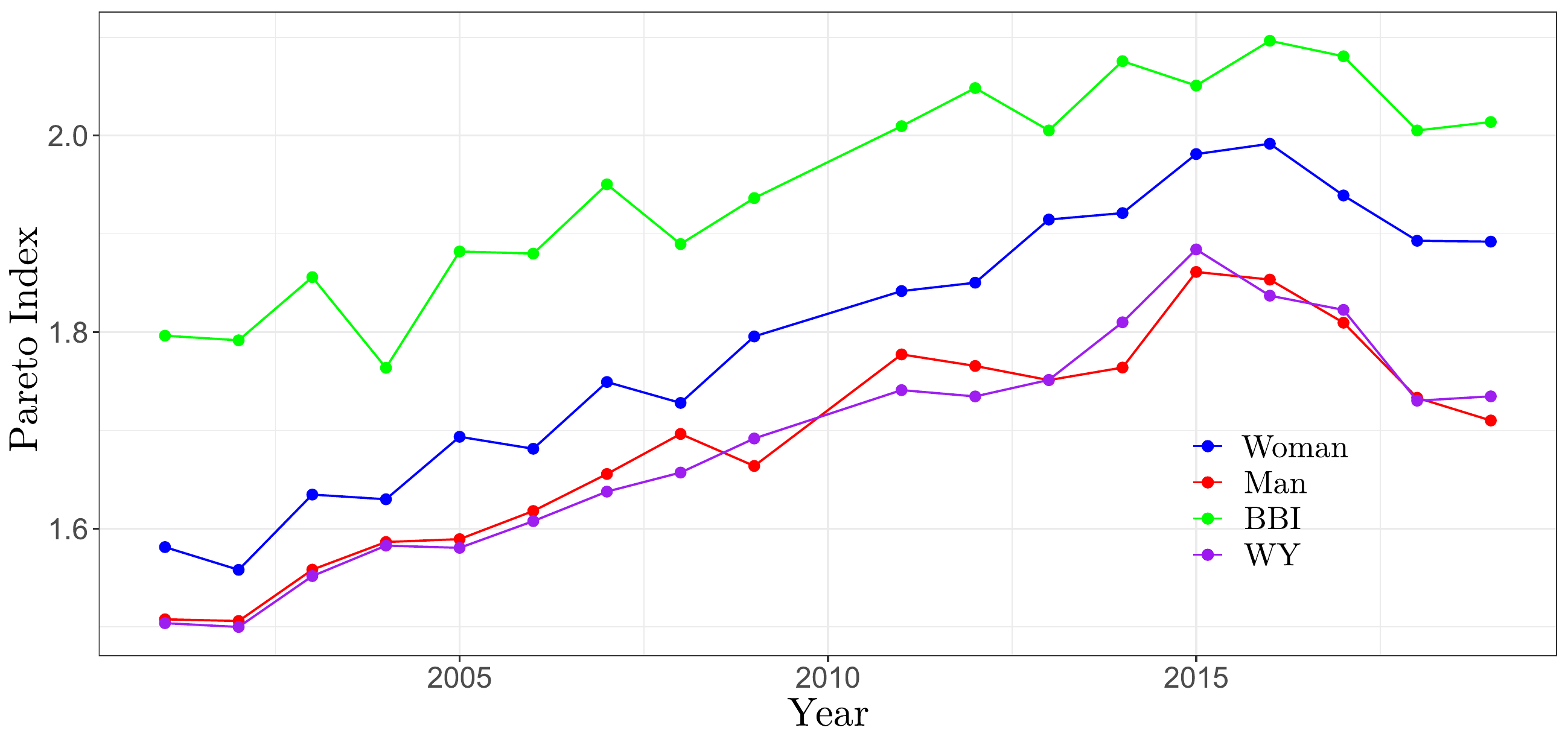}
        \caption{Pareto index series.}\label{fig:Spareto}%
    \end{subfigure}%
    ~ 
    \begin{subfigure}{0.5\linewidth}
        \includegraphics[width=\linewidth]{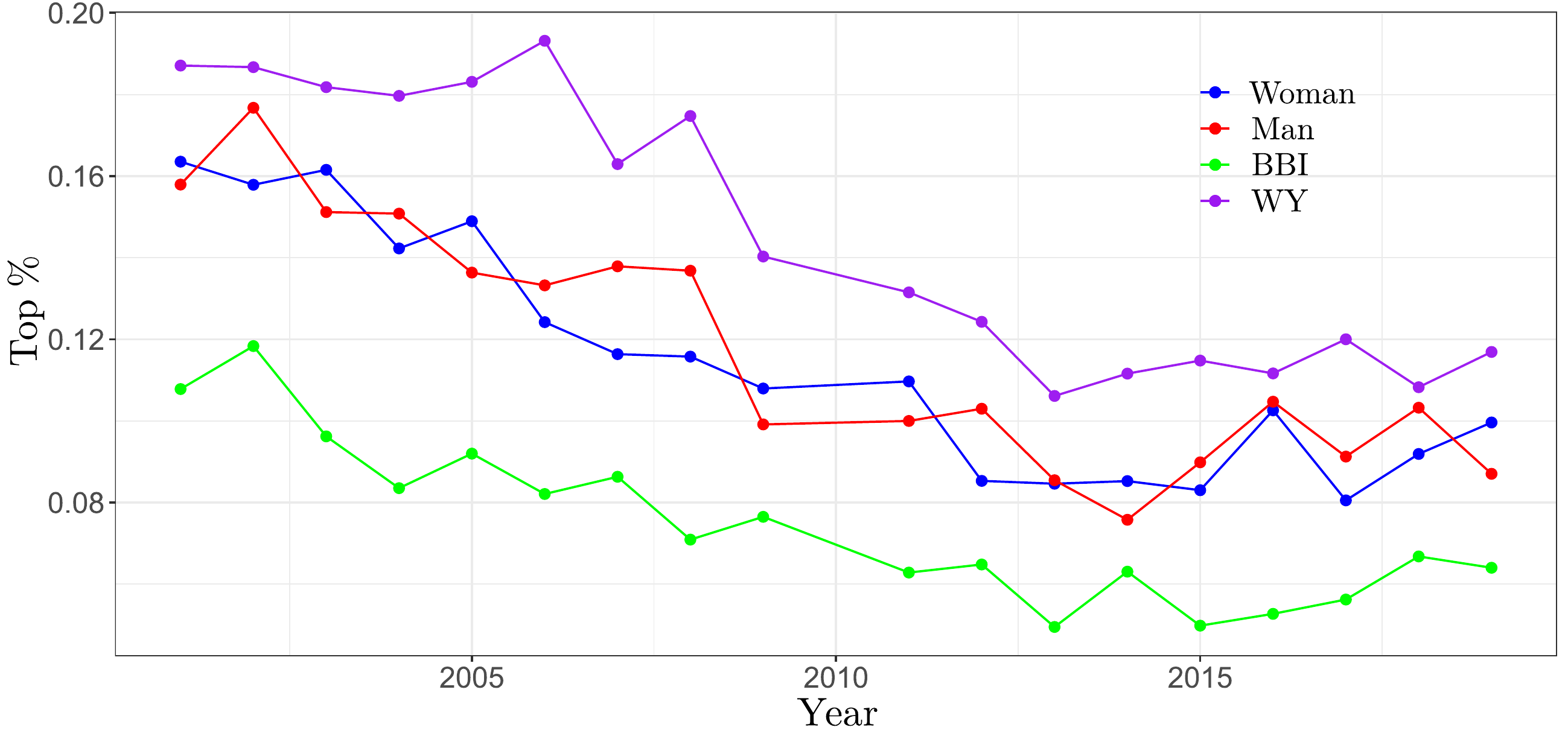}
        \caption{Top-percentage series.}\label{fig:Stop}
    \end{subfigure}

    \begin{subfigure}{0.5\linewidth}
        \includegraphics[width=\linewidth]{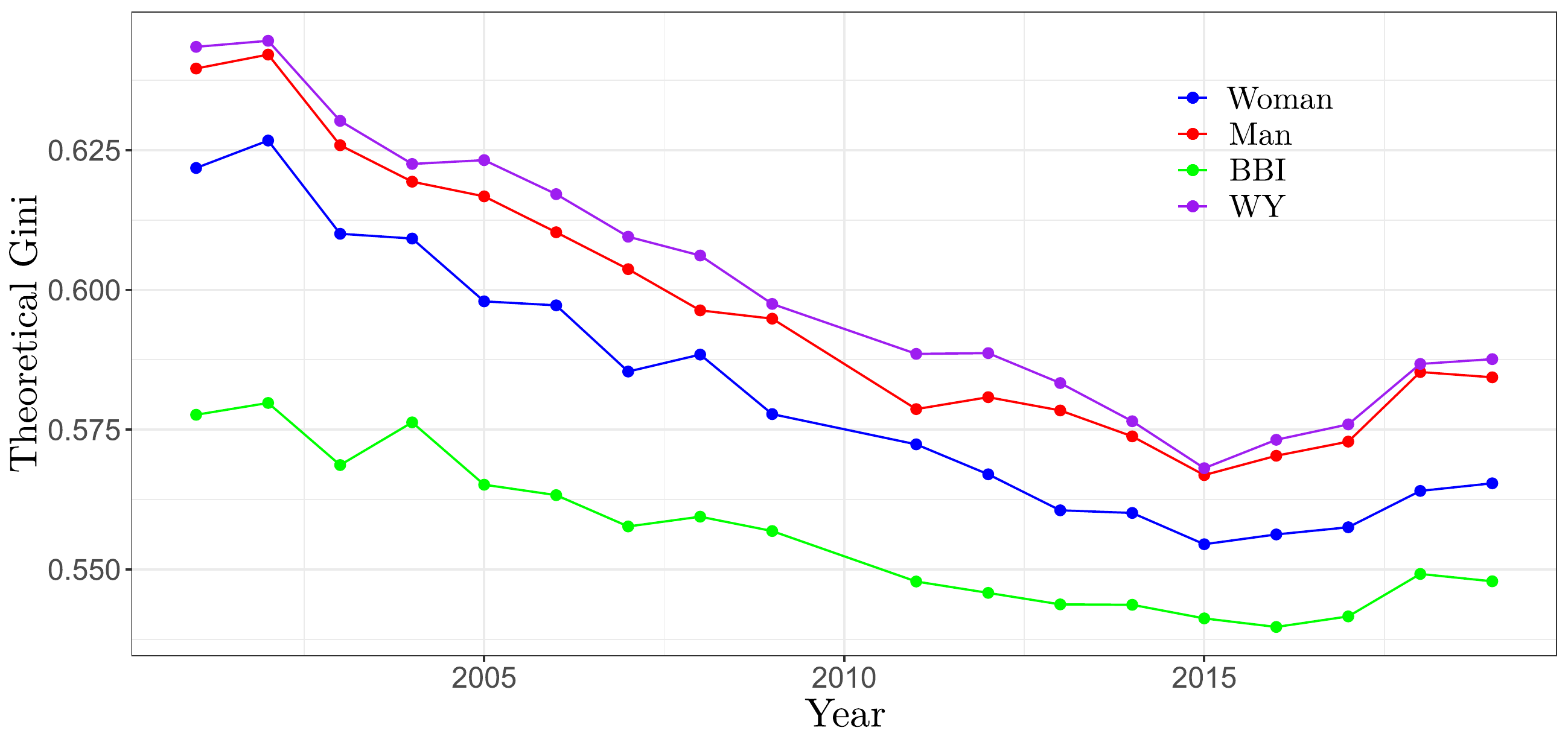}
        \caption{Theoretical Gini coefficient series, Eq. \ref{gini-t}.}\label{fig:STgini}%
    \end{subfigure}%
    ~ 
    \begin{subfigure}{0.5\linewidth}
        \includegraphics[width=\linewidth]{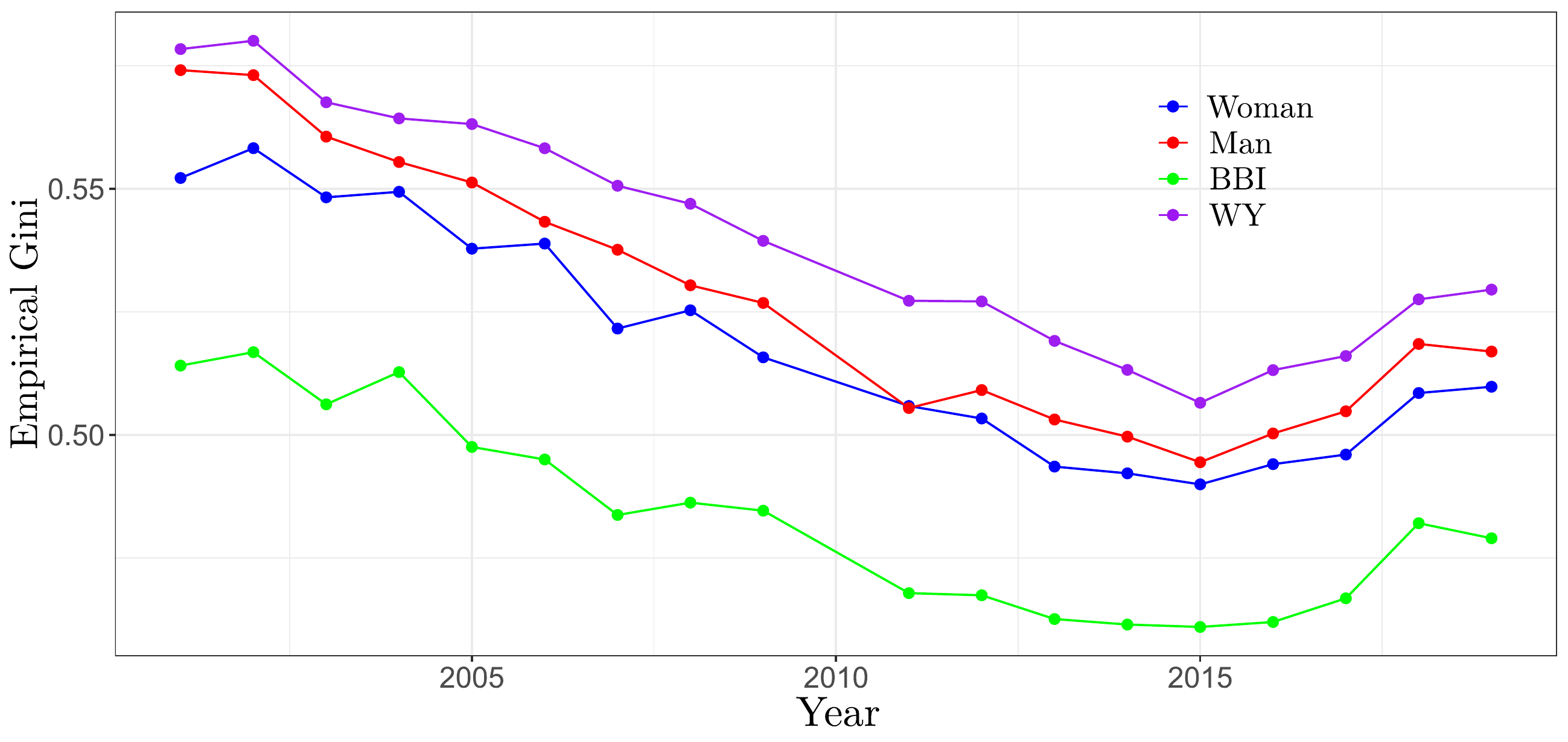}
        \caption{Empirical Gini coefficient series.}\label{fig:SEgini}
    \end{subfigure}
  
    \caption{Time series of the inequality indicators for the stratified data.}

\end{figure}

Like the previous section, we also did the same time series analysis with a stratified data using the two dichotomies described in that section. The stratified data time series has similar behavior between each subgroup. These time series of the subgroup data also present a similar behavior to the complete data set, as can be seen comparing the stratified data Figs. \cref{fig:ST,fig:Spareto,fig:Stop,fig:STgini,fig:SEgini} with the complete data Figs. \cref{fig:pareto,fig:top,fig:T,fig:gini}.

Now one can draw the same correlations using the stratified data. This will give us if these correlations are likely to be general or a specific correlation of the Brazilian time series data. The results of the Pearson correlation are shown in the Table \ref{tb:STtable}.

\begin{table}[H]
    \caption{\label{tb:STtable} Anticorrelation of Pareto index with top-percentage and correlation of the theoretical and empirical Gini coefficients for the stratified data.} 

     \begin{center}

        \begin{tabular}{lcccc}
        \hline 
         \multicolumn{5}{c}{Pearson Correlation} \\
        \cline{3-5} 
          &  & Pareto X Top \% &  & Empirical X Theoretical Gini \\
        \cline{3-3} \cline{5-5} 
         Man &  & $\displaystyle -0.858\pm 0.128$ &  & $\displaystyle 0.995\pm 0.025$ \\
        Woman &  & $\displaystyle -0.922\pm 0.097$$ $ &  & $\displaystyle 0.985\pm 0.043$ \\
        WY &  & $\displaystyle -0.902\pm 0.108\ $ &  & $\displaystyle 0.996\pm 0.024$ \\
        BBI &  & $\displaystyle -0.861\pm 0.127$ &  & $\displaystyle 0.983\pm 0.046$ \\
         \hline
        \end{tabular}
        \end{center}

\end{table}

According to our results, the Empirical and Theoretical Gini coefficients correlation is a really stable correlation, not varying much when switching the subgroup. The anti-correlation between the Pareto index and the top-percentage has more variance and, with exception of the woman subgroup, is weaker for the stratified data when compared to the original data.

\section{Conclusions}

The two class model is a well-tested hypothesis for the income distribution, being built-on around two famous distributions: the exponential BG and the Pareto power-law. It is important to remember that the exponential has stability in a multi-agent system, which the log-normal distribution lacks. 

Some previous studies have proposed the method to determine the crossover between the two distributions by using a log-log graph and manually trying to spot a discontinuity or a linear behavior. To our knowledge this paper provides for the first time a method to establish an optimal crossover income.

The optimal crossover method presented in this paper not only displays continuity, but also has a significantly lower RMSLE when comparing to a fixed proportion (5\%) for the Pareto region. The optimization was cross validated by a bootstrap out-of-the-bag method, which had a good performance in the test sets.

Analyzing stratified data and comparing the dichotomies revealed a greater inequality in the privileged groups (male/white and yellow) compared with their respective counterparts. The black/brown group exhibited the most equality and the least proportion participating in the Pareto region, having only 6.42\%.

Lastly, we analyse the temporal evolution of all indicators and draw two strong correlations. The first is the correlation between the theoretical Gini coefficient and the empirical Gini. The second is between the Top-percentage and Pareto index, which was found for the first time. These two correlations were also found in the stratified data, with the first having a strong correlation with low variance the second have an anti-correlation with more variance when we compare each subgroup. Further investigation in other countries is needed to generalize our findings using the Brazilian data.

The next step would be to implement this novel approach to other countries. Making simulations of a two-class model (define an empirical CCDF given a predetermined model), to determine a loss function that gives the best estimation of true value of the simulated model, is another step that would further validate an end-to-end method of fitting this model. The end goal would be to add a sample weighting and expansion, thus the distribution will have the correct sampling treatment. The sample weighting and expansion is a rather advanced topic of sampling statistics and usually dismissed in model regression.

%\bibliographystyle{unsrtnat}
%\bibliography{references}  %%% Uncomment this line and comment out the ``thebibliography'' section below to use the external .bib file (using bibtex) .
\clearpage

\medskip

\printbibliography

@article{Pareto,
  title     = {Cours d'{\'E}conomie Politique. By VILFREDO PARETO, Professeur {\`a} l'Universit{\'e} de Lausanne. Vol. I. Pp. 430. I896. Vol. II. Pp. 426. I897. Lausanne: F. Rouge},
  author    = {Moore, HL},
  journal   = {The ANNALS of the American Academy of Political and Social Science},
  volume    = {9},
  number    = {3},
  pages     = {128--131},
  year      = {1897},
  publisher = {Sage Publications Sage CA: Thousand Oaks, CA}
}

@article{hill,
  title     = {A simple general approach to inference about the tail of a distribution},
  author    = {Hill, Bruce M},
  journal   = {The annals of statistics},
  pages     = {1163--1174},
  year      = {1975},
  publisher = {JSTOR}
}

@article{EvIndia,
  title     = {The Pareto Law and the distribution of income},
  author    = {Shirras, G Findlay},
  journal   = {The Economic Journal},
  volume    = {45},
  number    = {180},
  pages     = {663--681},
  year      = {1935},
  publisher = {JSTOR}
}

@article{EvJap,
  title     = {Pareto's law for income of individuals and debt of bankrupt companies},
  author    = {Aoyama, Hideaki and Souma, Wataru and Nagahara, Yuichi and Okazaki, Mitsuhiro P and Takayasu, Hideki and Takayasu, Misako},
  journal   = {Fractals},
  volume    = {8},
  number    = {03},
  pages     = {293--300},
  year      = {2000},
  publisher = {World Scientific}
}

@article{EvUS,
  title     = {New evidence for the power-law distribution of wealth},
  author    = {Levy, Moshe and Solomon, Sorin},
  journal   = {Physica A: Statistical Mechanics and its Applications},
  volume    = {242},
  number    = {1-2},
  pages     = {90--94},
  year      = {1997},
  publisher = {Elsevier}
}

@incollection{Exp1,
  title     = {Pareto’s law of income distribution: Evidence for Germany, the United Kingdom, and the United States},
  author    = {Clementi, Fabio and Gallegati, Mauro},
  booktitle = {Econophysics of wealth distributions},
  pages     = {3--14},
  year      = {2005},
  publisher = {Springer}
}

@article{Exp2,
  title     = {Statistical mechanics of money},
  author    = {Dragulescu, Adrian and Yakovenko, Victor M},
  journal   = {The European Physical Journal B-Condensed Matter and Complex Systems},
  volume    = {17},
  number    = {4},
  pages     = {723--729},
  year      = {2000},
  publisher = {Springer}
}

@article{2US,
  title     = {Exponential and power-law probability distributions of wealth and income in the United Kingdom and the United States},
  author    = {Dr{\u{a}}gulescu, Adrian and Yakovenko, Victor M},
  journal   = {Physica A: Statistical Mechanics and its Applications},
  volume    = {299},
  number    = {1-2},
  pages     = {213--221},
  year      = {2001},
  publisher = {Elsevier}
}

@article{2RO,
  title     = {Income inequality in Romania: The \\ exponential-Pareto distribution},
  author    = {Oancea, Bogdan and Andrei, Tudorel and Pirjol, Dan},
  journal   = {Physica A: Statistical Mechanics and its Applications},
  volume    = {469},
  pages     = {486--498},
  year      = {2017},
  publisher = {Elsevier}
}

@article{Exp3,
  title     = {Temporal evolution of the "thermal" and "superthermal" income classes in the USA during 1983--2001},
  author    = {Silva, A Christian and Yakovenko, Victor M},
  journal   = {EPL (Europhysics Letters)},
  volume    = {69},
  number    = {2},
  pages     = {304},
  year      = {2004},
  publisher = {IOP Publishing}
}

@article{Opthreshold,
  title     = {Optimal threshold for Pareto tail modelling in the presence of outliers},
  author    = {Safari, Muhammad Aslam Mohd and Masseran, Nurulkamal and Ibrahim, Kamarulzaman},
  journal   = {Physica A: Statistical Mechanics and its Applications},
  volume    = {509},
  pages     = {169--180},
  year      = {2018},
  publisher = {Elsevier}
}

@inproceedings{Pso,
  title        = {Particle swarm optimization},
  author       = {Kennedy, James and Eberhart, Russell},
  booktitle    = {Proceedings of \\ ICNN'95-International Conference on Neural Networks},
  volume       = {4},
  pages        = {1942--1948},
  year         = {1995},
  organization = {IEEE}
}

@article{hybrid,
  title     = {A hybrid PSO-BFGS strategy for global optimization of multimodal functions},
  author    = {Li, Shutao and Tan, Mingkui and Tsang, Ivor W and Kwok, James Tin-Yau},
  journal   = {IEEE Transactions on Systems, Man, and Cybernetics, Part B (Cybernetics)},
  volume    = {41},
  number    = {4},
  pages     = {1003--1014},
  year      = {2011},
  publisher = {IEEE}
}

@article{2BR,
  title     = {Kinetic theory and Brazilian income distribution},
  author    = {Siciliani, Igor DS and Tragtenberg, Marcelo HR},
  journal   = {Physica A: Statistical Mechanics and its Applications},
  volume    = {513},
  pages     = {166--174},
  year      = {2019},
  publisher = {Elsevier}
}

@article{BFGSnon1,
  title     = {Limited memory BFGS for nonsmooth optimization},
  author    = {Skajaa, Anders},
  journal   = {Master's thesis},
  year      = {2010},
  publisher = {Citeseer}
}

@article{BFGSnon2,
  title     = {Nonsmooth variants of Powell's BFGS convergence theorem},
  author    = {Guo, Jiayi and Lewis, AS},
  journal   = {SIAM Journal on Optimization},
  volume    = {28},
  number    = {2},
  pages     = {1301--1311},
  year      = {2018},
  publisher = {SIAM}
}

@article{outboot,
  title     = {Estimating the error rate of a prediction rule: improvement on \\ cross-validation},
  author    = {Efron, Bradley},
  journal   = {Journal of the American statistical association},
  volume    = {78},
  number    = {382},
  pages     = {316--331},
  year      = {1983},
  publisher = {Taylor \& Francis}
}

@article{USt,
  title     = {Temporal evolution of the “thermal” and “superthermal” income classes in the USA during 1983--2001},
  author    = {Silva, A Christian and Yakovenko, Victor M},
  journal   = {EPL (Europhysics Letters)},
  volume    = {69},
  number    = {2},
  pages     = {304},
  year      = {2004},
  publisher = {IOP Publishing}
}

@incollection{JPt,
  title     = {Physics of personal income},
  author    = {Souma, Wataru},
  booktitle = {Empirical science of financial fluctuations},
  pages     = {343--352},
  year      = {2002},
  publisher = {Springer}
}

@inproceedings{IgorNon,
  title        = {A modified particle swarm optimizer},
  author       = {Shi, Yuhui and Eberhart, Russell},
  booktitle    = {1998 IEEE international conference on evolutionary computation proceedings. IEEE world congress on computational intelligence (Cat. No. 98TH8360)},
  pages        = {69--73},
  year         = {1998},
  organization = {IEEE}
}

@article{BootCV,
  title     = {Improvements on cross-validation: the 632+ bootstrap method},
  author    = {Efron, Bradley and Tibshirani, Robert},
  journal   = {Journal of the American Statistical Association},
  volume    = {92},
  number    = {438},
  pages     = {548--560},
  year      = {1997},
  publisher = {Taylor \& Francis}
}

@article{kalecki,
  title     = {On the Gibrat distribution},
  author    = {Kalecki, Micha{\l}},
  journal   = {Econometrica: Journal of the Econometric Society},
  pages     = {161--170},
  year      = {1945},
  publisher = {JSTOR}
}

@article{sim1,
  title     = {Pareto and Boltzmann--Gibbs behaviors in a deterministic multi-agent system},
  author    = {Gonzalez-Estevez, J and Cosenza, MG and Lopez-Ruiz, R and Sanchez, JR},
  journal   = {Physica A: Statistical Mechanics and its Applications},
  volume    = {387},
  number    = {18},
  pages     = {4637--4642},
  year      = {2008},
  publisher = {Elsevier}
}

@article{sim2,
  title     = {Pareto law in a kinetic model of market with random saving propensity},
  author    = {Chatterjee, Arnab and Chakrabarti, Bikas K and Manna, SS},
  journal   = {Physica A: Statistical Mechanics and its Applications},
  volume    = {335},
  number    = {1-2},
  pages     = {155--163},
  year      = {2004},
  publisher = {Elsevier}
}

@article{taleb,
  title     = {Gini estimation under infinite variance},
  author    = {Fontanari, Andrea and Taleb, Nassim Nicholas and Cirillo, Pasquale},
  journal   = {Physica A: Statistical Mechanics and its Applications},
  volume    = {502},
  pages     = {256--269},
  year      = {2018},
  publisher = {Elsevier}
}

%%% Uncomment this section and comment out the \bibliography{references} line above to use inline references.
% \begin{thebibliography}{1}

% 	\bibitem{kour2014real}
% 	George Kour and Raid Saabne.
% 	\newblock Real-time segmentation of on-line handwritten arabic script.
% 	\newblock In {\em Frontiers in Handwriting Recognition (ICFHR), 2014 14th
% 			International Conference on}, pages 417--422. IEEE, 2014.

% 	\bibitem{kour2014fast}
% 	George Kour and Raid Saabne.
% 	\newblock Fast classification of handwritten on-line arabic characters.
% 	\newblock In {\em Soft Computing and Pattern Recognition (SoCPaR), 2014 6th
% 			International Conference of}, pages 312--318. IEEE, 2014.

% 	\bibitem{hadash2018estimate}
% 	Guy Hadash, Einat Kermany, Boaz Carmeli, Ofer Lavi, George Kour, and Alon
% 	Jacovi.
% 	\newblock Estimate and replace: A novel approach to integrating deep neural
% 	networks with existing applications.
% 	\newblock {\em arXiv preprint arXiv:1804.09028}, 2018.

% \end{thebibliography}

\end{document}